\documentclass[10pt,twocolumn,letterpaper]{article}

\usepackage{times}
\usepackage{epsfig}
\usepackage{graphicx}
\usepackage{amsmath,amssymb}
\usepackage{booktabs}
\usepackage{authblk}
\usepackage{url}
\usepackage{hyperref}
\usepackage{xcolor}
\usepackage{caption}

\hypersetup{
    colorlinks=true,
    linkcolor=blue,
    urlcolor=blue,
    citecolor=blue,
}

\newcommand{\sysName}{SPLICE}

\title{\sysName: Part-Level 3D Shape Editing \\
from Local Semantic Extraction to Global Neural Mixing}

\author[1]{Jin Zhou}
\author[1]{Hongliang Yang}
\author[1]{Pengfei Xu\thanks{Corresponding author: xupengfei.cg@gmail.com}}
\author[1]{Hui Huang}

\affil[1]{CSSE, Shenzhen University}

\date{}

\begin{document}
\twocolumn[{
    \maketitle
    \vspace{-1.5em}
    
    \begin{center}
        \includegraphics[width=0.9\linewidth]{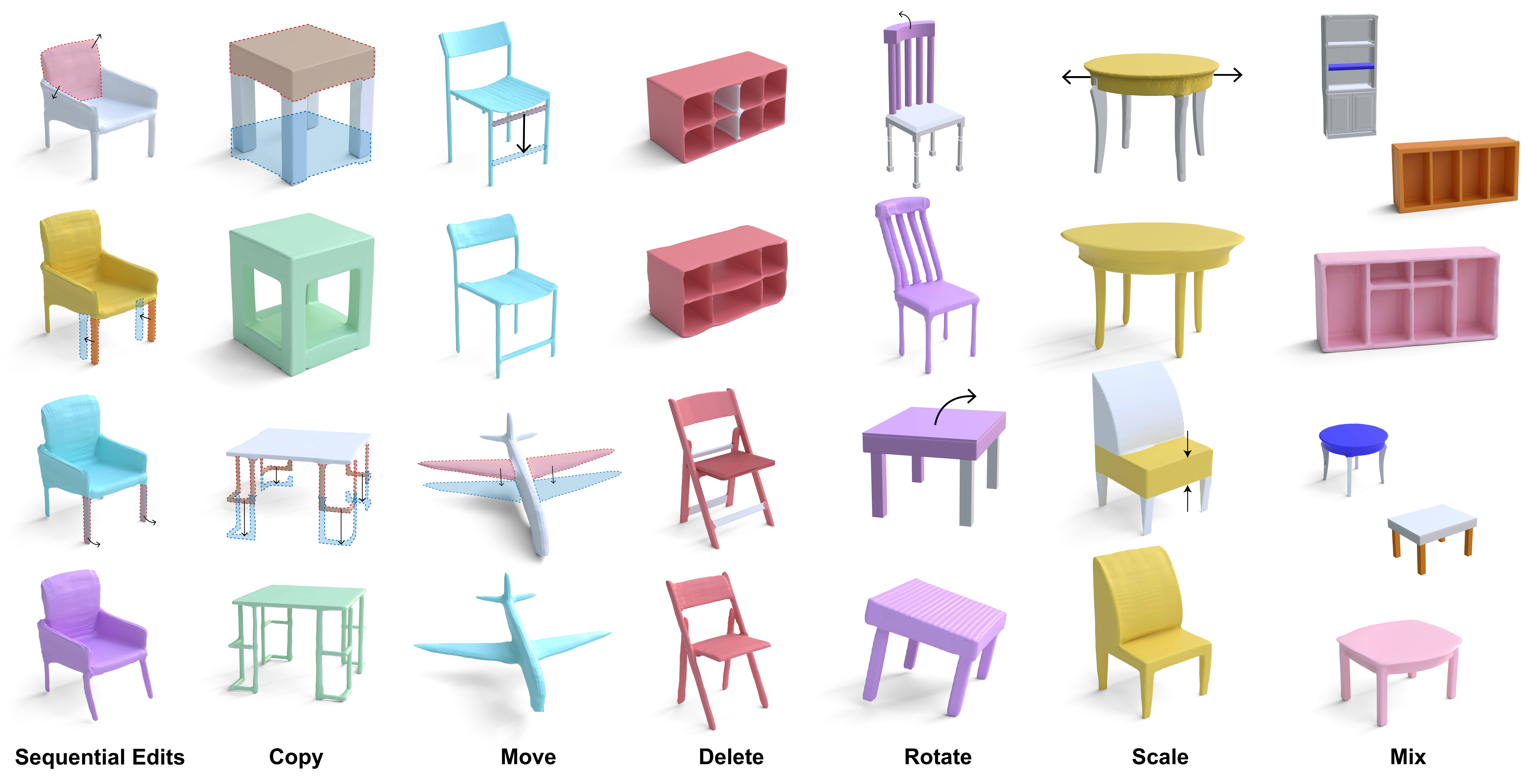}
        
        \vspace{0.5em} 
        
        \captionof{figure}{Part-level editing results produced by \sysName. Our method supports a wide range of intuitive editing operations, including sequential edits, copy, move, delete, rotate, scale, and mix, without requiring manual post-adjustment. The edited shapes remain structurally coherent and visually plausible. Additionally, \sysName\ exhibits strong robustness under multi-step editing, consistently maintaining high reconstruction quality throughout the editing process.}
        \label{fig:teaser}
    \end{center}
    
    \vspace{1.0em}
}]


\begin{abstract}
Neural implicit representations of 3D shapes have shown great potential in 3D shape editing due to their ability to model high-level semantics and continuous geometric representations. However, existing methods often suffer from limited editability, lack of part-level control, and unnatural results when modifying or rearranging shape parts. In this work, we present \sysName, a novel part-level neural implicit representation of 3D shapes that enables intuitive, structure-aware, and high-fidelity shape editing. By encoding each shape part independently and positioning them using parameterized Gaussian ellipsoids, \sysName\ effectively isolates part-specific features while discarding global context that may hinder flexible manipulation. A global attention-based decoder is then employed to integrate parts coherently, further enhanced by an attention-guiding filtering mechanism that prevents information leakage across symmetric or adjacent components. Through this architecture, \sysName\ supports various part-level editing operations, including translation, rotation, scaling, deletion, duplication, and cross-shape part mixing. These operations enable users to flexibly explore design variations while preserving semantic consistency and maintaining structural plausibility. Extensive experiments demonstrate that \sysName\ outperforms existing approaches both qualitatively and quantitatively across a diverse set of shape-editing tasks.
\end{abstract}

\section{Introduction} \label{sec:intro}
3D shape editing is a longstanding research topic in computer graphics. Recently, neural implicit representations of 3D shapes have emerged as powerful tools for modeling 3D data, gaining increasing attention for their application in shape editing tasks~\cite{kusupati2024semantic, liu2021deepmetahandles, hertz2022spaghetti, hu2024cns, hao2020dualsdf, berzins23bs, lyu2023controllable}. These representations encode shapes as continuous functions, offering strong expressive power and learnability. They can effectively capture high-level semantic attributes of shapes, such as part connectivity, spatial arrangement, and global symmetry, making them promising candidates for producing natural and semantically consistent editing outcomes.

However, current neural implicit shape editing methods face several limitations. Methods based on latent space optimization aim to find an alternative code in the latent space that satisfies user constraints~\cite{hu2024cns, lyu2023controllable}, but they do not guarantee the preservation of geometric details from the original shape. Moreover, the lack of interpretability in the latent space often results in unintuitive or uncontrollable editing trajectories. On the other hand, part-level or proxy-based editing methods have achieved success in deformation-style edits using abstractions like control points~\cite{liu2021deepmetahandles, hu2024cns}, Gaussian components~\cite{hertz2022spaghetti, koo2023salad}, or sphere-based proxies~\cite{hao2020dualsdf}. Nevertheless, these methods struggle with more complex tasks such as part relocation or recombination. Moved parts often remain influenced by their original context, failing to form new plausible connections in their updated positions, which leads to unnatural deformations that compromise overall shape quality~\cite{berzins23bs}.

To address these challenges, we propose \sysName\ (\textbf{S}hape \textbf{P}art-\textbf{L}evel neural \textbf{I}mplicit representation via Lo\textbf{C}al Semantic \textbf{E}xtraction), a novel representation designed to support flexible and semantically aware editing. Our method treats shape parts as the basic editing units, providing three key advantages: high editing freedom, strong preservation of local features, and improved structural coherence.
The core idea lies in decoupling each part's representation from the global context, ensuring that it is reconstructed solely based on its own geometry. To achieve this, we first independently encode each part using a convolutional encoder, extracting both its geometry and pose in the form of latent shape codes and Gaussian ellipsoid parameters. Rather than using raw ellipsoid coefficients, we represent poses via six spatially distributed ellipsoid vertices, which more effectively couple position, orientation, and scale. During training, we simulate user edits via random affine transformations and encode the perturbed pose using a SIREN network~\cite{sitzmann2020implicit}. These operations help the model learn to treat pose as an editable, disentangled input. In the decoding stage, we employ a Transformer-based occupancy decoder that enables each query point to attend to part embeddings. To prevent unwanted information leakage between symmetric or neighboring components, we further introduce an attention-guiding loss that encourages points to favor their corresponding part. Finally, we optionally incorporate a latent diffusion model to support automatic shape refinement and part completion when edits introduce large-scale inconsistencies or missing components.

In summary, \sysName\ offers four core innovations: (1) part-level latent code separation for geometry-pose disentanglement, (2) ellipsoid-vertex-based pose encoding that preserves editable spatial semantics, (3) an attention-guiding loss in a Transformer decoder to reduce inter-part interference, and (4) a diffusion-based refinement pipeline to improve editing robustness and enable part synthesis. Together, these innovations enable \sysName\ to support more natural, controllable, and semantically consistent shape manipulation. Extensive experiments confirm that our method outperforms prior approaches in both reconstruction fidelity and editing flexibility.


\begin{figure*}
    \centering
    \includegraphics[width=\linewidth]{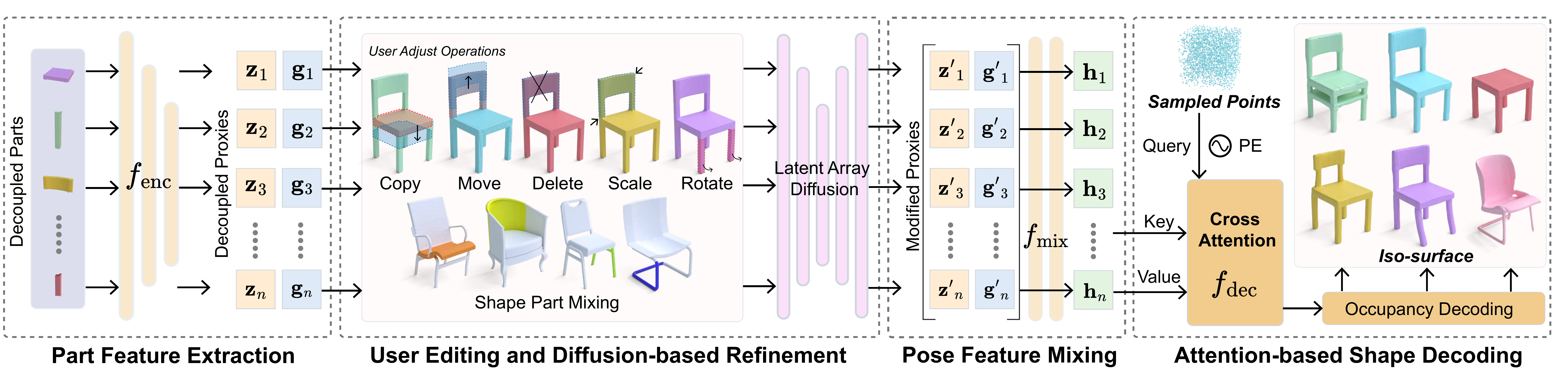}
    \caption{Overview of our \sysName\ pipeline. Given a 3D shape decomposed into parts, we first apply \textbf{Part Feature Extraction} using a shared convolutional encoder $f_{\mathrm{enc}}$ to obtain per-part geometry latent codes $\{\mathbf{z}_i\}$ and Gaussian proxies $\{\mathbf{g}_i\}$. In \textbf{User Editing and Diffusion-based Refinement}, these proxies can be directly modified by user operations (e.g., move, scale, mix) or adjusted by a latent diffusion model $f_{\mathrm{adj}}$ to restore global coherence. The resulting updated proxies $\{(\mathbf{z}'_i, \mathbf{g}'_i)\}$ are then processed in \textbf{Pose Feature Mixing}, where each $\mathbf{g}'_i$ is encoded by a SIREN-based pose encoder $\phi$, and combined with $\mathbf{z}'_i$ via a multilayer perceptron $f_{\mathrm{MLP}}$ to obtain the final part embedding $\mathbf{h}_i$. Finally, in \textbf{Attention-based Shape Decoding}, sampled query points attend to $\{\mathbf{h}_i\}$ through a cross-attention transformer $f_{\mathrm{dec}}$, followed by occupancy decoding to reconstruct the final shape via marching cubes.}
    \label{fig:pipeline}
\end{figure*}

\section{Related Work} \label{sec:related}

\subsection{Neural Implicit Representations}
Neural implicit representations of 3D shapes have seen rapid advances in 3D reconstruction over recent years. Methods such as Occupancy Networks~\cite{mescheder2019occupancy,Zhang20233DShape2VecSet}, IM-NET~\cite{chen2019learning}, DeepSDF~\cite{park2019deepsdf}, and NeRF~\cite{Mildenhall2020NeRF} have demonstrated impressive capabilities to represent complex topology and fine-grained details using continuous shape fields parameterized by global latent codes or network weights. Despite their representational strengths, these global latent representations inherently lack explicit geometric primitives for direct user manipulation. As a result, interactive editing remains cumbersome, usually necessitating latent optimization or retraining rather than intuitive, local adjustments. This limitation motivates our pursuit of an implicit editing framework that combines learned field fidelity with interpretable, localized editability.

\subsection{Segmented Implicit Representations}
A parallel line of research decomposes implicit shapes into localized, independently decoded components, which are subsequently merged to reconstruct the full shape. Methods such as BSP-Net~\cite{Chen2019BSPNetGC}, Neural-Patch Learning~\cite{Paschalidou2021NeuralPL}, articulated part models~\cite{Sun2019LearningAH}, primitive-based reconstruction~\cite{tulsiani2017learning}, and early volumetric component models~\cite{Funkhouser2004ModelingBE} facilitate responsive local edits but typically sacrifice seamless integration between parts and global shape coherence. Specifically, these approaches often treat parts as fully independent during encoding and decoding without mechanisms to enforce consistency or repair shape continuity after editing. As a result, edited shapes may contain disconnected or mismatched regions, and achieving structural plausibility often requires manual post-processing. To improve consistency, Xu et al.~\cite{xu2023unsupervised} propose a part-retrieval-and-assembly pipeline that reconstructs 3D shapes by retrieving and composing local implicit components, improving reconstruction quality through explicit segmentation priors. Chen et al.~\cite{chen2024dae} develop DAE-Net, a deforming auto-encoder that jointly performs fine-grained shape co-segmentation and deformation, yielding consistently aligned implicit part decompositions and enabling precise local adjustments. To further address the integration gap, boundary coalescing techniques~\cite{yin2020coalesce} and hybrid methods that blend local and global features~\cite{lin2022neuform} have emerged, though they still struggle to maintain high-quality junctions and holistic shape integrity. Our method, in contrast, leverages a unified implicit encoding that directly references individual proxies, ensuring coherent global reconstruction alongside local editability.

\subsection{Latent Space Shape Editing}
Another prevalent approach to implicit shape editing focuses on directly navigating or manipulating the latent space. Methods such as CNS-Edit~\cite{hu2024cns}, SHAP-EDITOR~\cite{chen2024shap} treat editing as latent-space traversal or optimization, mapping user-defined operations (e.g., translation, scaling, removal) to movements within global latent representations. Variants introduced by Zheng et al.~\cite{zheng2023locally}, Kusupati et al.~\cite{kusupati2024semantic}, and Berzins et al.~\cite{berzins23bs} accelerate editing through feed-forward latent predictions, enabling near-real-time edits. Other notable contributions, including Controllable Editing (SLIDE)~\cite{lyu2023controllable}, DeepMetaHandles~\cite{liu2021deepmetahandles}, and DualSDF~\cite{hao2020dualsdf}, extend this latent manipulation paradigm by exploring disentanglement and hierarchical latent structures. However, such latent-based methods inherently constrain edits to the learned latent manifold, limiting editability to representable shape variations and complicating the identification of precise latent trajectories for desired edits. Consequently, they frequently cause unintended global shape modifications, leading to a loss of identity and detail fidelity.
In contrast, our method avoids reliance on a fixed latent space of whole shapes, enabling edits beyond the original learned distribution while explicitly maintaining shape identity and detail preservation.

\subsection{Compositional Implicit Shape Editing}
Recent trends in graphics and vision communities highlight the value of compositional implicit shape models, where complex shapes are represented as collections of simpler, semantically meaningful sub-parts. Early efforts, such as DualSDF~\cite{hao2020dualsdf}, partition shapes into hierarchical SDF layers, facilitating coarse-to-fine editing yet lacking precise local control. DeepMetaHandles~\cite{liu2021deepmetahandles} utilize learned handle proxies but retain global shape priors that limit edit flexibility. Wei et al.~\cite{wei2020learning} introduce a semantic-parameter inference framework that predicts interpretable deformation controls for implicit shapes, enabling intuitive part edits via high-level parameters. Li et al.~\cite{li2022editvae} present EditVAE, an unsupervised parts-aware VAE that decomposes point-cloud shapes into controllable part-specific latent codes, facilitating part generation and manipulation without explicit supervision. SPAGHETTI~\cite{hertz2022spaghetti} introduced unsupervised implicit segment disentanglement, allowing intuitive part-level manipulation. Further developments like PartNeRF~\cite{Tertikas2023GeneratingPE}, Implicit Neural Head Synthesis~\cite{chen2023implicit}, and SALAD~\cite{koo2023salad} improve part-specific rendering and deformation capabilities, though challenges persist in seamless integration and boundary clarity. 3DShape2VecSet~\cite{Zhang20233DShape2VecSet} proposed a flexible latent-vector-based representation, enhancing local fidelity but lacking strong semantic control. SLIDE~\cite{lyu2023controllable} and Semantic Implicit Templates~\cite{kusupati2024semantic} demonstrate precise local adjustments via point-based proxies and parametric implicit surfaces yet remain globally coupled by fitting constraints.
Although existing compositional methods offer clear benefits, most still rely on global decompositions that do not fully isolate each component. As a result, unintended inter-part interactions can occur and accumulate over successive edits, ultimately degrading the shape quality. In contrast, our framework encodes each editable part independently and represents its pose via sampled ellipsoid vertices. We then apply random rigid perturbations during training and introduce attention-guiding decoding to prevent cross-part leakage. This combination ensures that local geometric edits remain precise and effective while the overall shape stays coherent and structurally consistent.

\section{Method} \label{sec:method}

We aim to construct a flexible and part-aware 3D shape representation that supports intuitive and precise editing. To this end, we model a shape as a set of independently encoded parts, where each part captures only its own geometry and pose, explicitly disentangled from the global context. This local encoding ensures that user operations, such as translating, rotating, scaling, deleting, duplicating, or mixing parts, can be performed in a direct and interpretable way while maintaining semantic consistency across the final shape.

Our pipeline consists of four main stages, as illustrated in Figure~\ref{fig:pipeline}. First, in \textbf{Part Feature Extraction}, we decompose a shape into parts and use a 3D convolutional encoder $f_{\mathrm{enc}}$ to extract a geometry code $\mathbf{z}_i$ and a pose proxy $\mathbf{g}_i$ for each part. Instead of relying on raw ellipsoid parameters, we represent the pose of each part using six strategically sampled surface points from each Gaussian proxy, encouraging the model to reason about position, orientation, and scale jointly.
Second, in \textbf{User Editing and Diffusion-based Refinement}, users can directly modify the pose proxy of a part $\mathbf{g}_i$ to perform local edits and optionally invoke a latent diffusion model to adjust part geometry and pose in a globally coherent way. This dual mechanism supports both fine-grained control and high-level structural consistency.
Third, in \textbf{Pose Feature Mixing}, we encode the edited pose using a SIREN-based~\cite{sitzmann2020implicit} network $\phi$ and fuse it with the updated geometry code $\mathbf{z}'_i$ via a lightweight MLP. Notably, we avoid the early use of self-attention here to preserve part-level independence before decoding.
Finally, in \textbf{Attention-based Shape Decoding}, a Transformer-based decoder $f_{\mathrm{dec}}$ uses cross-attention to condition each query point on relevant part embeddings. To prevent ambiguity, especially in symmetric or densely packed structures, we introduce an attention-guiding loss that encourages each point to attend to its true source part, thus preserving semantic locality during reconstruction.

\subsection{Part Feature Extraction}

This stage aims to extract part-specific shape features $\mathbf{z}_i$ and initial pose embeddings $\mathbf{g}_i$ from shape data annotated with part labels. To accomplish this, we utilize datasets that include explicit part segmentations, such as PartNet~\cite{mo2019partnet}, or employ existing shape segmentation algorithms~\cite{chen2024dae, chen2019bae} on datasets without segmentations, such as ShapeNet~\cite{Chang2015ShapeNetAI}. Since available segmentations are typically annotated on shape surfaces rather than shape voxels, we determine the part label of each voxel in shape by associating it with the nearest annotated surface point.
Each identified part is then centered and encoded independently as a sparse voxel grid $\mathbf{V}_i$. We utilize an encoder function $f_{\mathrm{enc}}$, composed of alternating convolution and pooling layers, to generate the compact shape feature vector $\mathbf{z}_i$ and an initial Gaussian embedding $\tilde{\mathbf{g}}_i$:
\begin{equation}
(\mathbf{z}_i, \tilde{\mathbf{g}}_i) = f_{\mathrm{enc}}(\mathbf{V}_i).
\end{equation}
where $(\mathbf{z}_i, \tilde{\mathbf{g}}_i)$ is predicted by the encoder, from which we split and interpret the parameters as follows:
\begin{align}
\boldsymbol{\mu}_i, \boldsymbol{\sigma}_i &= \mathrm{split}(\mathbf{z}_i), \\
\mathbf{t}_i, \mathbf{s}_i, \mathbf{q}_i &= \mathrm{split}(\tilde{\mathbf{g}}_i),
\end{align}
where $\boldsymbol{\mu}_i$ and $\boldsymbol{\sigma}_i$ are the mean and standard deviation of the part’s latent shape code, $\mathbf{t}_i$ is a local translation offset, $\mathbf{s}_i$ is the scale vector, and $\mathbf{q}_i$ is the quaternion representing orientation.
The full Gaussian parameterization $\mathbf{g}_i$ is then computed as:
\begin{equation}
\mathbf{g}_i = (\mathbf{c}_i + \mathbf{t}_i,\ \boldsymbol{\lambda}_i \cdot \mathbf{s}_i,\ \mathbf{R}(\mathbf{q}_i)),
\end{equation}
where $\mathbf{c}_i = [c_x, c_y, c_z]$ is the centroid of the voxel grid, $\boldsymbol{\lambda}_i = [\lambda_1, \lambda_2, \lambda_3]$ represents global scaling factors, and $\mathbf{R}(\mathbf{q}_i)$ converts the quaternion into a rotation matrix.

Although one could compute Gaussian parameters for each part through PCA on its voxel coordinates rather than predicting them via the network, this PCA-based approach presents two key limitations. First, PCA cannot learn a consistent mapping between pose representation and shape features; it treats orientation and scale as purely geometric properties without capturing semantic correlations. Second, the eigenvector directions obtained from PCA exhibit ambiguity: for symmetric parts such as cylinders, the two minor principal axes are interchangeable, leading to unstable or unpredictable orientation estimates. In contrast, by learning Gaussian parameters jointly with shape features, our model ensures that pose and geometry remain semantically and numerically consistent.

Because our encoder directly predicts these Gaussian parameters, we need a way to measure and improve how well they fit the actual voxel distribution of each part. To this end, we introduce a negative log-likelihood loss that explicitly supervises the predicted Gaussian against the part’s voxels:
\begin{equation}
\mathcal{L}_{\mathrm{nll}} = - \frac{1}{n} \sum_{j=1}^n \log \mathcal{N}(\mathbf{x}_j \mid \mathbf{g}_i).
\end{equation}
The part-level latent code is then regularized with a KL-divergence term against a unit Gaussian prior:
\begin{equation}
\mathcal{L}_{\mathrm{KL}} = \frac{1}{2} \sum_{k=1}^d \left( \mu_{i,k}^2 + \sigma_{i,k}^2 - \log \sigma_{i,k}^2 - 1 \right).
\end{equation}


\subsection{User Editing and Diffusion-based Refinement}

With our method, user edits can operate directly on each part’s Gaussian parameters. For example, to edit a shape part, dragging is achieved by adjusting the position offset $\mathbf{t}_i$, rotating is achieved by updating the quaternion $\mathbf{q}_i$, and scaling is achieved by modifying $\mathbf{s}_i$. Because these edits occur before part integration, the decoder treats them as new facts, enabling precise and localized control. In many cases, this direct manipulation suffices, and users can immediately render the edited shape without further processing.

For effective editing, the model should strictly adhere to the new Gaussian parameters rather than relying on its implicit positional assumptions. Indeed, even when voxel grids of each part are centered, a model might infer approximate original positions from learned biases. To mitigate this issue, we adopt a strategy inspired by Hertz et al.~\cite{hertz2022spaghetti}.
Specifically, during training, we sample random affine transformations to simulate user editing operations. These transformations are simultaneously applied to Gaussian parameters and the query points sampled from target shapes. The transformations include rotation, translation, and scaling, ensuring shape integrity without introducing distortions.

However, purely local edits can become tedious when the edits are extreme and neglect global context. Manually ensuring that every part remains consistent with the overall structure can be cumbersome. Moreover, a local edit may produce parts whose geometry no longer seamlessly aligns with the edited pose, leading to subtle mismatches or gaps. To address these issues, we introduce an optional latent diffusion model inspired by Koo et al.~\cite{koo2023salad}, which users may invoke whenever they desire global refinement or missing-part synthesis. 

Our diffusion model refines part poses and latent shape codes in a unified manner. Each part is represented by a concatenated feature vector including its validity flag, pose embedding, and shape code, which is progressively denoised toward a coherent configuration. This joint refinement gradually leads to semantically plausible global arrangements and geometrically consistent parts, as pose and shape are coordinated throughout the denoising process. At inference time, the model can synthesize entirely new parts to fill holes, adjust mismatched geometry, or complete partially edited shapes. While invoking diffusion may slightly sacrifice some fine-grained detail, it enforces global structural coherence and reduces manual effort.

Formally, given an initial part embedding $(\mathbf{z}_i, \mathbf{g}_i)$, we define the adjustment function $f_{\mathrm{adj}}$, which may represent either a user edit or a diffusion-based refinement, to produce an updated pair:
\begin{equation}
(\mathbf{z}'_i, \mathbf{g}'_i) = f_{\mathrm{adj}}(\mathbf{z}_i, \mathbf{g}_i),
\end{equation}
where $\mathbf{g}'_i$ reflects the modified pose, and $\mathbf{z}'_i$ is optionally refined to ensure compatibility with the new configuration.
Since the diffusion refinement is optional, users can choose to apply it only when necessary. If a simple part-level edit already satisfies design intent, the system can bypass the diffusion refinement and render the result directly. Otherwise, users can invoke the diffusion model to ensure that all parts integrate naturally into a cohesive shape.

\subsection{Pose Feature Mixing}

In this stage, our goal is to compute the final part embedding $\mathbf{h}_i$ by combining geometry and pose information from each edited or refined part. Formally, this corresponds to learning a mixing function $f_{\mathrm{mix}}$ that transforms the updated shape code $\mathbf{z}'_i$ and pose proxy $\mathbf{g}'_i$ into a unified representation:
\begin{equation}
\mathbf{h}_i = f_{\mathrm{mix}}(\mathbf{z}'_i, \mathbf{g}'_i).
\end{equation}

To represent the pose information effectively, we extract six representative 3D points from the updated Gaussian ellipsoid $\mathbf{g}'_i$. These points correspond to the endpoints of the three principal axes of the ellipsoid, sampled symmetrically in both directions. Concretely, we take the center of the ellipsoid and extend it along each principal direction by a fixed radius of 1.75, forming six vertices: $\{\mathbf{v}_{i,j}\}_{j=1}^6$. These points are intended to lie near the outer surface of the part, providing a compact yet expressive representation that encodes position, orientation, and scale jointly.

We choose six vertices instead of directly inputting raw ellipsoid parameters to tightly couple positional, rotational, and scale attributes, forcing the network to consider all pose information jointly. Even if parts are centered, models can otherwise encode rotation and scale into the latent shape code and ignore explicit pose signals. By encoding six spatially distributed vertices along the ellipsoid’s principal axes, the network is encouraged to consider all components of the pose (position, orientation, and scale) equally. This balanced representation helps prevent the model from ignoring or collapsing pose information during learning. We demonstrate the impact of this design in our ablation study.
Our method concatenates these six 3D points into a single 18-dimensional vector:
\begin{equation}
\mathbf{v}_i^{\mathrm{concat}} = \bigl[\mathbf{v}_{i,1};\ \mathbf{v}_{i,2};\ \dots;\ \mathbf{v}_{i,6}\bigr] \in \mathbb{R}^{18}.
\end{equation}
This concatenated vector is then encoded by a SIREN network~\cite{sitzmann2020implicit} to produce the part pose feature:
\begin{equation}
\mathbf{p}_i = \phi\bigl(\mathbf{v}_i^{\mathrm{concat}}\bigr)\in\mathbb{R}^D,
\end{equation}
where $\mathbf{p}_i$ is the part’s pose feature with dimension $D$. We then concatenate it with the updated part shape feature $\mathbf{z}'_i$
and feed the combined vector into a multilayer perceptron $\mathrm{MLP}$ to obtain the final part embedding $\mathbf{h}_i$:
\begin{equation}
\mathbf{h}_i = \mathrm{MLP}\bigl([\mathbf{z}'_i;\mathbf{p}_i]\bigr).
\end{equation}
Notably, unlike SPAGHETTI~\cite{hertz2022spaghetti},
we do not employ transformer-based encodings at this stage to avoid unintended information leakage between parts. Issues stemming from such information leakage and the effectiveness of our strategy will be discussed and demonstrated thoroughly in Section~\\ref{sec:Ablation}. 

\subsection{Attention-based Shape Decoding}

We represent each shape as an occupancy field, where the model predicts the probability of occupancy at a queried point $\mathbf{x} \in \mathbb{R}^3$:
\begin{equation}
    \hat{o}(\mathbf{x}) = f_{\mathrm{dec}}(\mathbf{x}; \{\mathbf{h}_i\}_{i=1}^N),
\end{equation}
where $f_{\mathrm{dec}}$ denotes the decoding function and $\{\mathbf{h}_i\}$ are the encoded part features. A high-resolution $128^3$ voxel grid is queried and the resulting field is converted into an explicit mesh using the marching cubes algorithm~\cite{lorensen1998marching}.

To decode occupancy values for each point $\mathbf{x}$, we adopt a Transformer-based structure with four layers and eight attention heads per layer. The point coordinate $\mathbf{x}$ is encoded via a SIREN-based Fourier mapping $\phi(\mathbf{x})$ and used to attend over the part embeddings $\{\mathbf{h}_i\}$, producing a locally conditioned feature $\mathbf{h}_x$:
\begin{equation}
\mathbf{h}_x = \mathrm{Transformer}(\phi(\mathbf{x}), \{\mathbf{h}_i\}).
\end{equation}
We then concatenate $\phi(\mathbf{x})$ and $\mathbf{h}_x$ and decode the occupancy through a 4-layer MLP:
\begin{equation}
\hat{o}(\mathbf{x}) = \mathrm{MLP}([\phi(\mathbf{x}); \mathbf{h}_x]).
\end{equation}
Although attending to all part embeddings can improve reconstruction quality, it may also introduce undesirable information leakage, especially in symmetric structures. For example, during training, a point near the left arm of a chair might attend to the right arm, causing unnatural deformations if only one side is edited. Moreover, when parts are extremely close, such as multiple slats in a chair back, unguided attention can fail to distinguish them, leading the decoder to conflate adjacent parts and produce blurred, fused outputs.
To mitigate this, we introduce an attention-guiding loss that encourages each internal point to attend primarily to its corresponding part. During training, we identify the ground-truth part for each sample point $\mathbf{x}$ and apply a soft cross-entropy loss over the attention weights:
\begin{equation}
    \mathcal{L}_{\mathrm{attn}} = - \sum_{i=1}^N \mathbf{w}_x^{(i)} \log \alpha_x^{(i)},
\end{equation}
where $\alpha_x^{(i)}$ is the attention weight from the query point $\mathbf{x}$ to part $i$, and $\mathbf{w}_x^{(i)}$ is a one-hot ground-truth part label. This loss is down-weighted to prevent the attention mechanism from collapsing into a nearest-neighbor lookup while still nudging the model to prefer its designated part in ambiguous cases.
The predicted occupancy $o(\mathbf{x})$ is supervised with a binary cross-entropy loss against the ground-truth occupancy $\hat{o}(\mathbf{x})$:
\begin{equation}
\mathcal{L}_{\mathrm{occ}} \;=\; -\,o(\mathbf{x})\,\log \hat{o}(\mathbf{x}) \;-\;\bigl(1 - o(\mathbf{x})\bigr)\,\log\bigl(1 - \hat{o}(\mathbf{x})\bigr).
\end{equation}

\textbf{Overall Loss.} The full objective combines the occupancy loss, Gaussian negative log-likelihood, KL divergence for latent shape codes, and the attention-guiding regularization:
\begin{equation}
    \mathcal{L}_{\mathrm{total}} = \mathcal{L}_{\mathrm{occ}} + \lambda_{\mathrm{nll}} \mathcal{L}_{\mathrm{nll}} + \lambda_{\mathrm{KL}} \mathcal{L}_{\mathrm{KL}} + \lambda_{\mathrm{attn}} \mathcal{L}_{\mathrm{attn}},
\end{equation}
where $\lambda_{\mathrm{nll}}$, $\lambda_{\mathrm{KL}}$, and $\lambda_{\mathrm{attn}}$ control the contribution of each term.
\section{Experiments}

\subsection{Datasets}
We conduct experiments on two large-scale shape datasets, i.e., PartNet~\cite{mo2019partnet} and ShapeNet~\cite{Chang2015ShapeNetAI}. PartNet provides fine-grained segmentation annotations across multiple object categories. Therefore, we can directly apply our method to this dataset. On the contrary, ShapeNet does not possess such part annotations. To apply our method, we first apply DAE-Net~\cite{chen2024dae} to segment shapes from ShapeNet~\cite{Chang2015ShapeNetAI} into semantic parts, ensuring consistent part-level supervision.

\subsection{Implementation Details}
\textbf{Voxel Representation.} In our method, each part is converted into a binary voxel grid of resolution $256^3$. The voxel is centered on the part’s centroid before being fed into the encoder. 

\textbf{Network Architectures.} In Part Feature Extraction, the convolutional encoder consists of successive $3\times3\times3$ convolutional layers with pooling, downsampling the input from $256^3$ to $1^3$, yielding a 256-dimensional shape code $\mathbf{z}_i$ and a 256-dimensional preliminary Gaussian embedding $\tilde{\mathbf{g}}_i$. In Pose Feature Mixing, we encode the concatenated 18-dimensional Gaussian vertices with a SIREN network to obtain a 128-dimensional pose feature $\mathbf{p}_i$. This is concatenated with the 256-dimensional shape feature and passed through a multi-layer perceptron ($\mathrm{MLP}$) to produce a final 256-dimensional part embedding $\mathbf{h}_i$.

\textbf{Transformer and Decoding.} For Attention-based Shape Decoding, each query point $\mathbf{x}$ is encoded to a 128-dimensional vector via the same SIREN mapping. We use a 4-layer Transformer with 8 attention heads per layer to compute a 128-dimensional local feature $\mathbf{h}_x$. A subsequent 4-layer $\mathrm{MLP}$ decodes $[\mathbf{h}_x;\phi(\mathbf{x})]$ to a scalar occupancy probability.

\textbf{Diffusion-based Refinement.} For the refinement module, we employ an 8-layer Transformer with 16 attention heads. Each input part is represented as a concatenation of three components: a binary validity flag (indicating whether the part is present),the Gaussian embedding $\mathbf{g}_i$,and the shape code $\mathbf{z}_i$, yielding a total dimension of $1+15+256$. This vector is first processed by an $\mathrm{MLP}$ to project it into a 512-dimensional space, and the network predicts the injected noise for denoising. To stabilize training, we also normalize and rescale the different input components so that their magnitudes remain balanced.

\textbf{Training Details.} We train all models using the Adam optimizer with a learning rate of $1\times10^{-4}$. The batch size is set to 40 shapes per iteration. Loss weights are chosen as $\lambda_{\mathrm{nll}} = 1$, $\lambda_{\mathrm{KL}} = 1\mathrm{e}{-5}$, and $\lambda_{\mathrm{attn}} = 0.1$. All models converge after approximately one day of training on 8 NVIDIA RTX 3090 GPUs.

\subsection{Baselines}

We benchmark our approach against two leading part-level editing frameworks. SPAGHETTI~\cite{hertz2022spaghetti} represents shapes as collections of Gaussian primitives, allowing intuitive local edits by adjusting primitive parameters. DualSDF~\cite{hao2020dualsdf} uses dual signed distance fields to capture both coarse and fine geometry, supporting smooth deformations via implicit field optimization under user-defined constraints.
We also considered CNS-Edit~\cite{hu2024cns}, which guides implicit voxel-field editing through a sequence of optimization steps and shares similar goals to our work. However, since its implementation is not publicly available, we exclude it from our quantitative and qualitative comparisons.
\subsection{Reconstruction Comparison}

\begin{table}[t]
\centering
\caption{Reconstruction quality comparison on the Chair and Airplane categories. ``w/o attention'' denotes our variant without attention guidance.}
\label{tab:recon}
\begin{tabular}{lccc}
\toprule
\multicolumn{4}{c}{\textbf{Chair}} \\
\midrule
Method & CD $\downarrow$ & EMD $\downarrow$ & Mesh ACC $\downarrow$ \\
\midrule
DualSDF~\cite{hao2020dualsdf}                  & 0.337           & 11.467          & 20.051 \\
SPAGHETTI~\cite{hertz2022spaghetti}            & \textbf{0.261}  & \textbf{9.660}  & \textbf{17.677} \\
Ours (w/ attention)                            & \underline{0.279}           & \underline{10.780}         & \underline{18.423} \\
Ours (w/o attention)                           & 0.349           & 11.130          & 21.480 \\
\midrule
\multicolumn{4}{c}{\textbf{Airplane}} \\
\midrule
Method & CD $\downarrow$ & EMD $\downarrow$ & Mesh ACC $\downarrow$ \\
\midrule
DualSDF~\cite{hao2020dualsdf}                  & 0.280           & 10.586          & \underline{7.606}  \\
SPAGHETTI~\cite{hertz2022spaghetti}            & \underline{0.090}           & \textbf{7.046}  & 8.587  \\
Ours (w/ attention)                            & \textbf{0.052}  & \underline{7.120}           & \textbf{7.423}  \\
Ours (w/o attention)                           & 0.458           & 11.432          & 34.592 \\
\bottomrule
\end{tabular}
\end{table}

Before evaluating editing capabilities, we first verify that our introduced representation for editability preserves reconstruction fidelity. We compare our method with SPAGHETTI, DualSDF, and a variant of our method without attention guidance. Experiments are conducted on the Chair and Airplane categories using identical data splits for all methods to ensure fairness.

As reported in Table~\ref{tab:recon}, our full model consistently matches or outperforms baseline methods in reconstruction quality. Notably, in the Airplane category, our method achieves the best overall performance, benefiting from the stable part decompositions produced by DAE-Net~\cite{chen2024dae}. This result highlights the robustness of our method across both manual annotations and automatically generated segmentations.

The comparison between our full model and the variant without attention guidance further underscores the effectiveness of our attention supervision strategy. By encouraging each query point to attend primarily to its corresponding part features, the model achieves better reconstruction fidelity, especially in challenging cases involving symmetry or closely packed parts.

\subsection{Qualitative Comparison}

\begin{figure}[t]
    \centering
    \includegraphics[width=\linewidth]{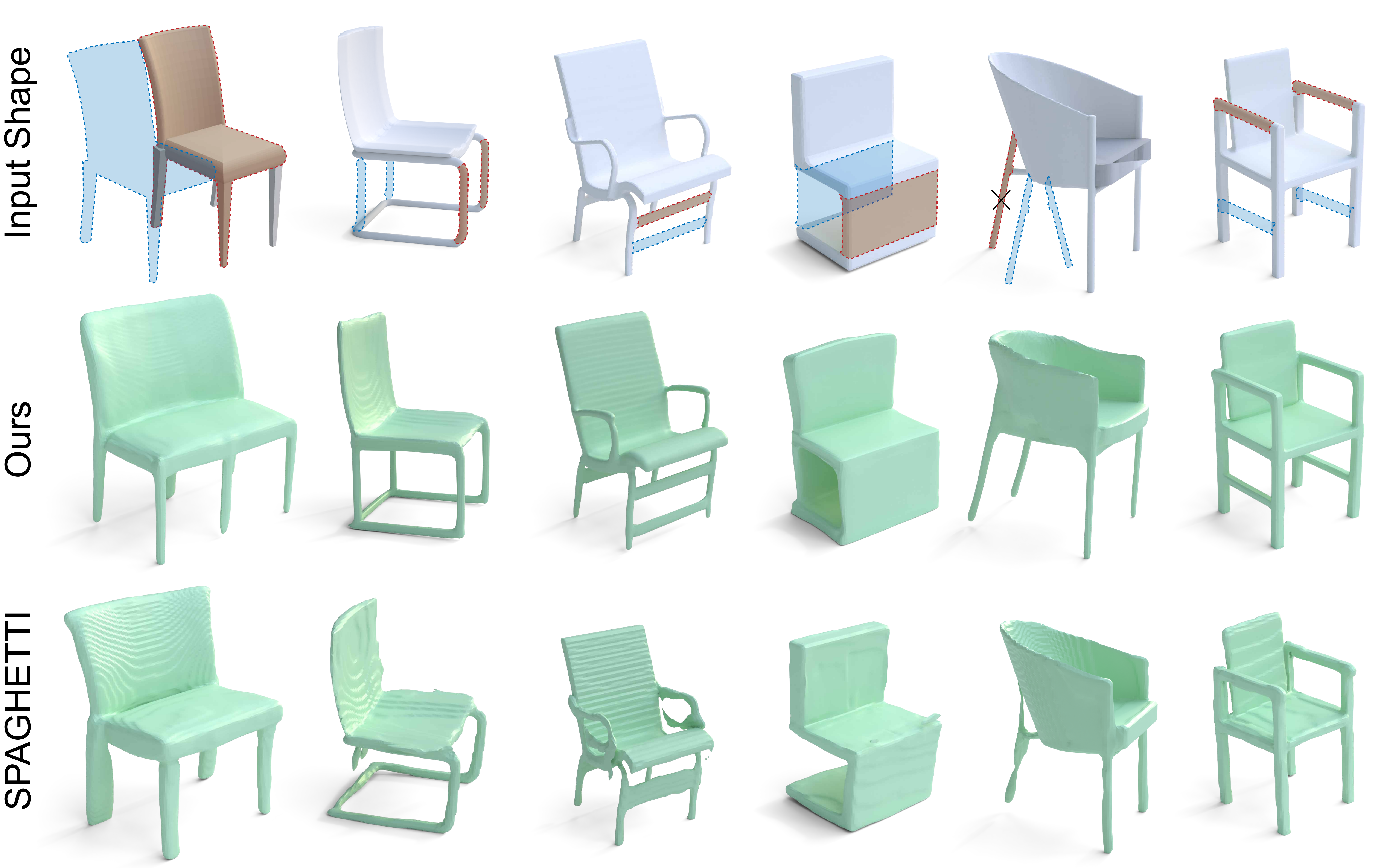}
    \caption{Comparison of the copy editing results between our method and SPAGHETTI. 
    }
    \label{fig:exp-op-copy}
\end{figure}

\begin{figure}[t]
    \centering
    \includegraphics[width=\linewidth]{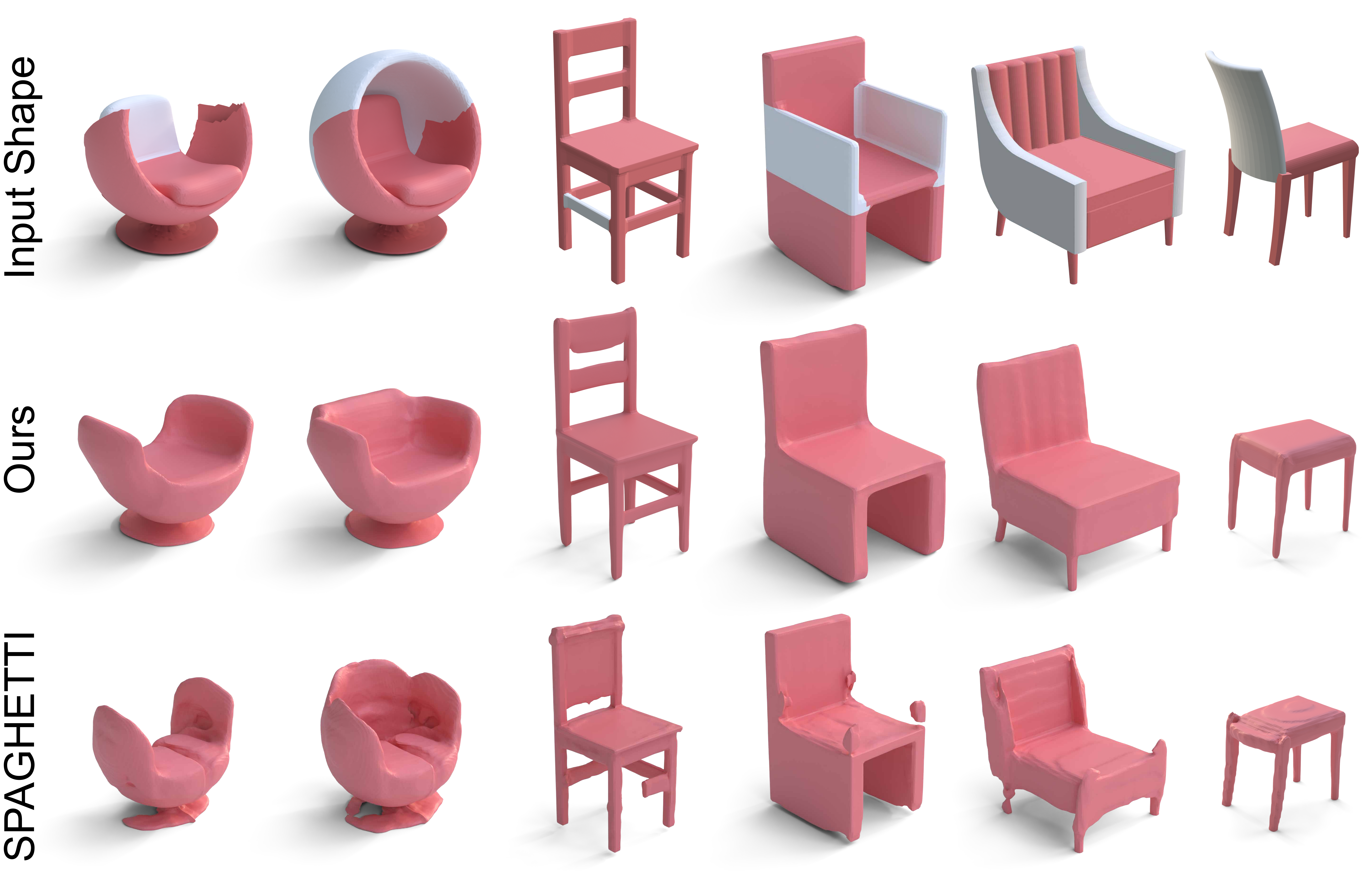}
    \caption{Comparison of delete editing results between our method and SPAGHETTI. The \textbf{gray} parts are deleted during the editing.
    }
    \label{fig:exp-op-delete}
\end{figure}

\begin{figure}[t]
    \centering
    \includegraphics[width=\linewidth]{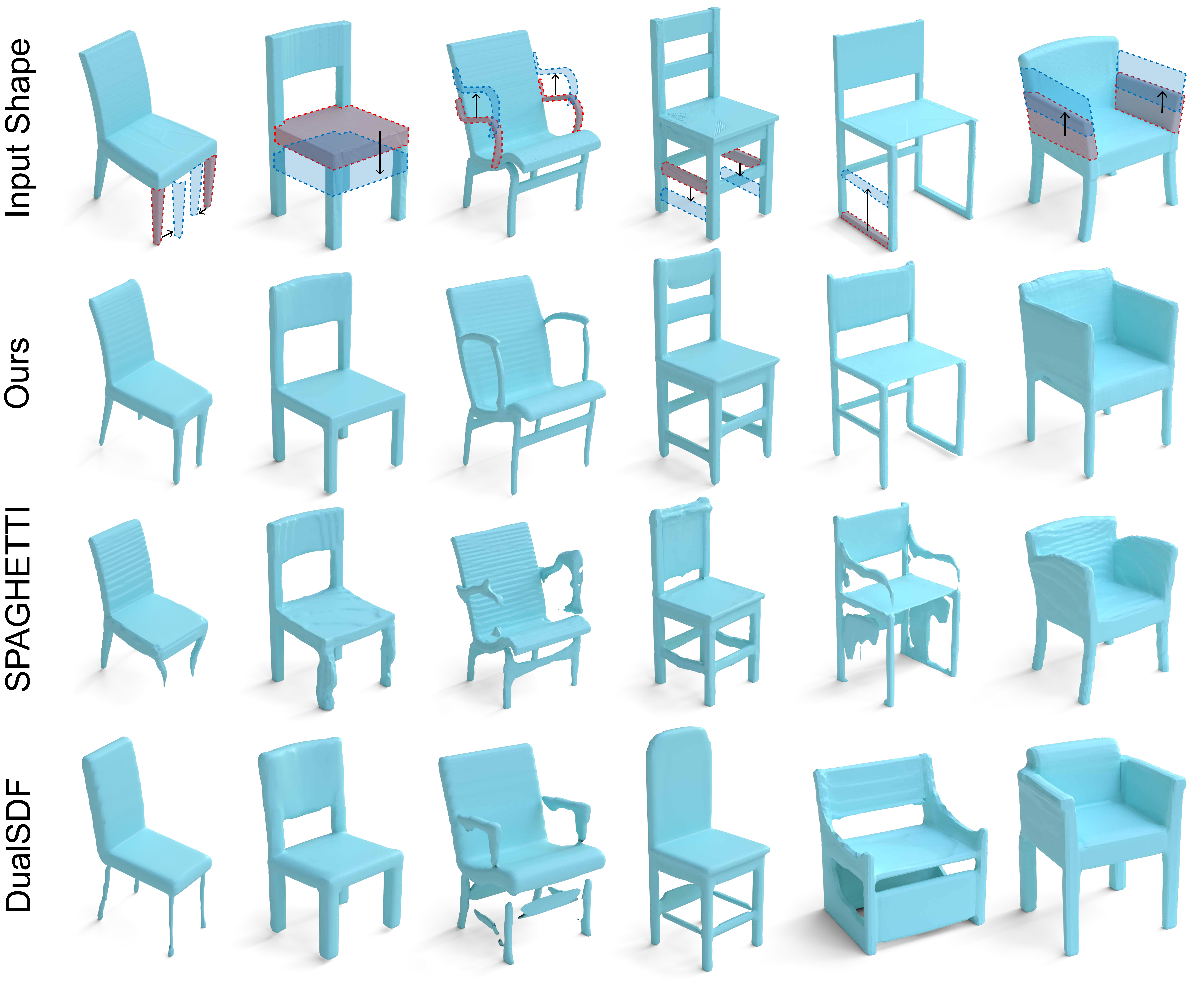}
    \caption{Comparison of move editing results between our method and SPAGHETTI and DualSDF.
    }
    \label{fig:exp-op-move}
\end{figure}

\begin{figure}[t]
    \centering
    \includegraphics[width=\linewidth]{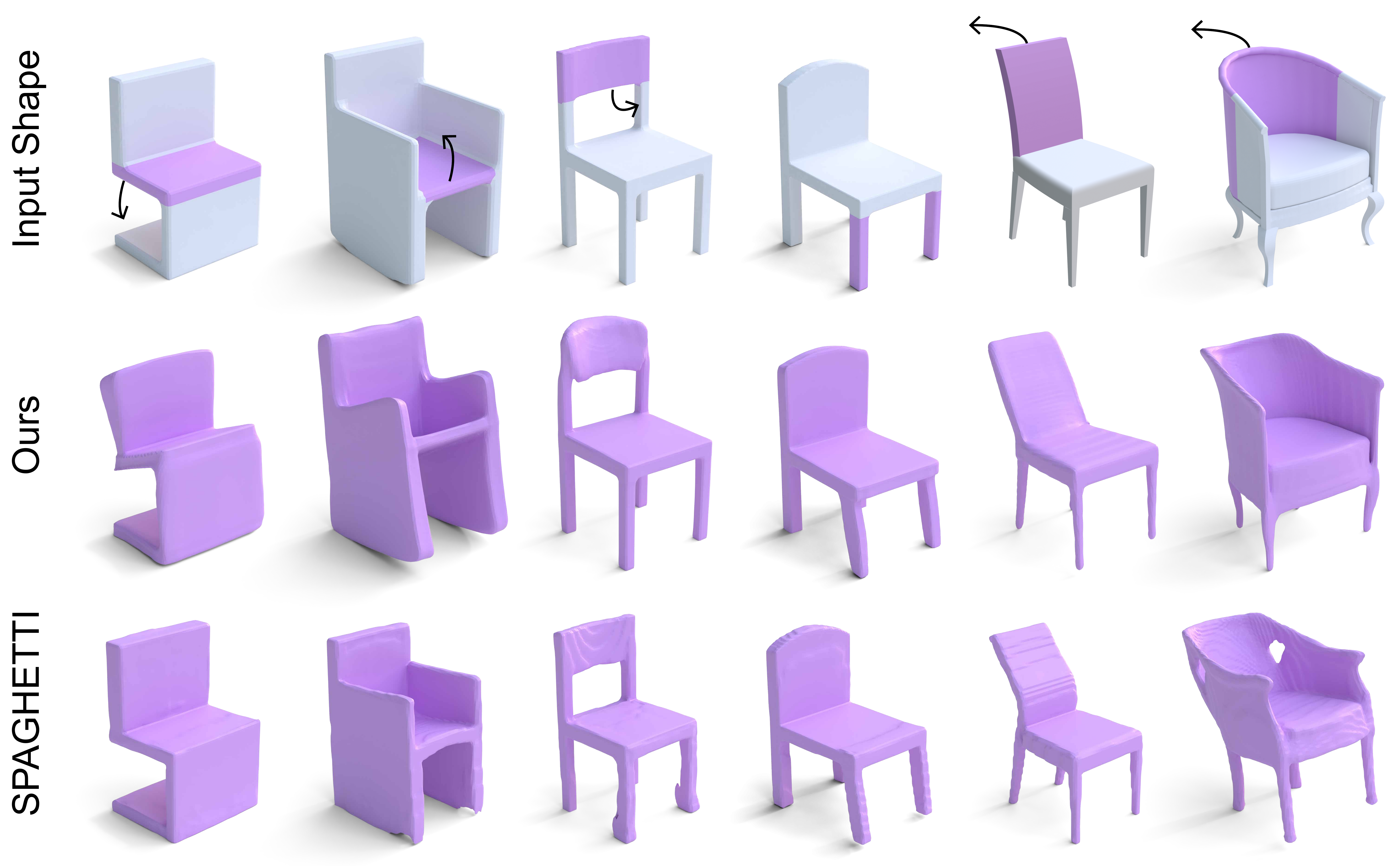}
    \caption{Comparison of rotation editing results between our method and SPAGHETTI. The \textbf{purple} parts are rotated during the editing.
    }
    \label{fig:exp-op-rotate}
\end{figure}

\begin{figure}[t]
    \centering
    \includegraphics[width=\linewidth]{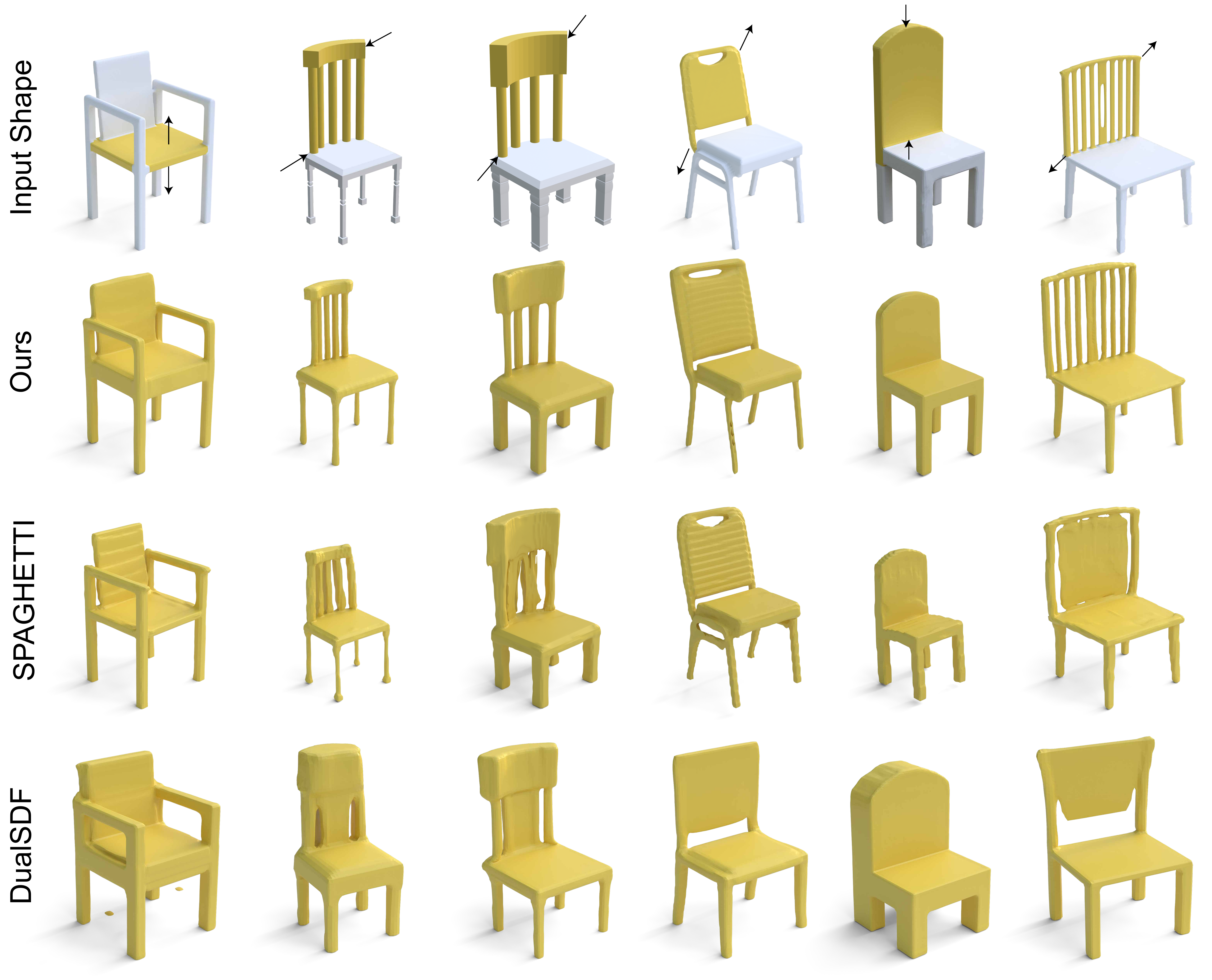}
    \caption{Comparison of scaling editing results between our method, SPAGHETTI, and DualSDF. The \textbf{yellow} parts are scaled during the editing.
    }
    \label{fig:exp-op-scale}
\end{figure}

\begin{figure}[t]
    \centering
    \includegraphics[width=\linewidth]{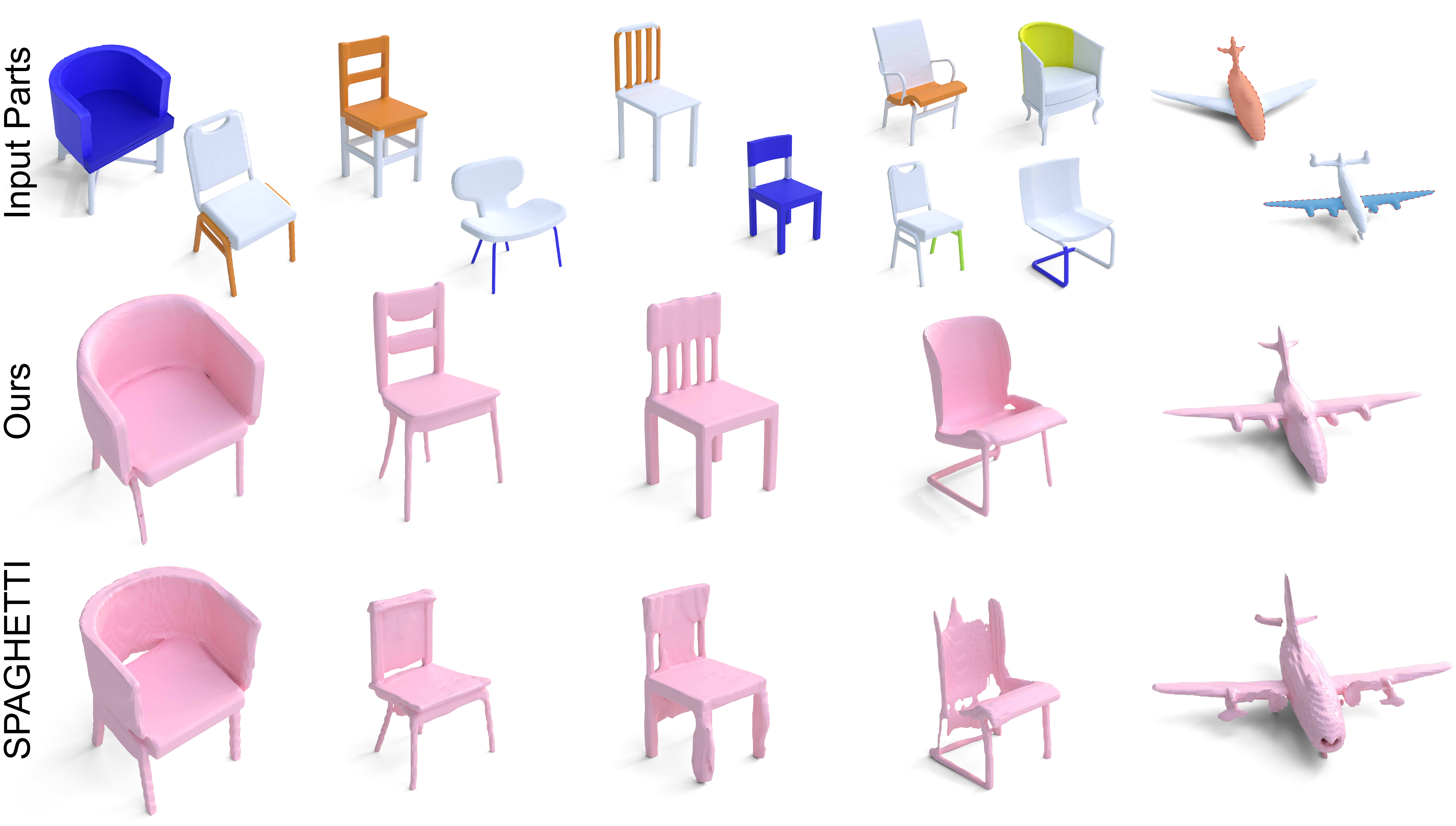}
    \caption{Comparison of part mixing results between our method and SPAGHETTI. 
    }
    \label{fig:exp-op-mix}
\end{figure}

\begin{figure}[t]
    \centering
    \includegraphics[width=\linewidth]{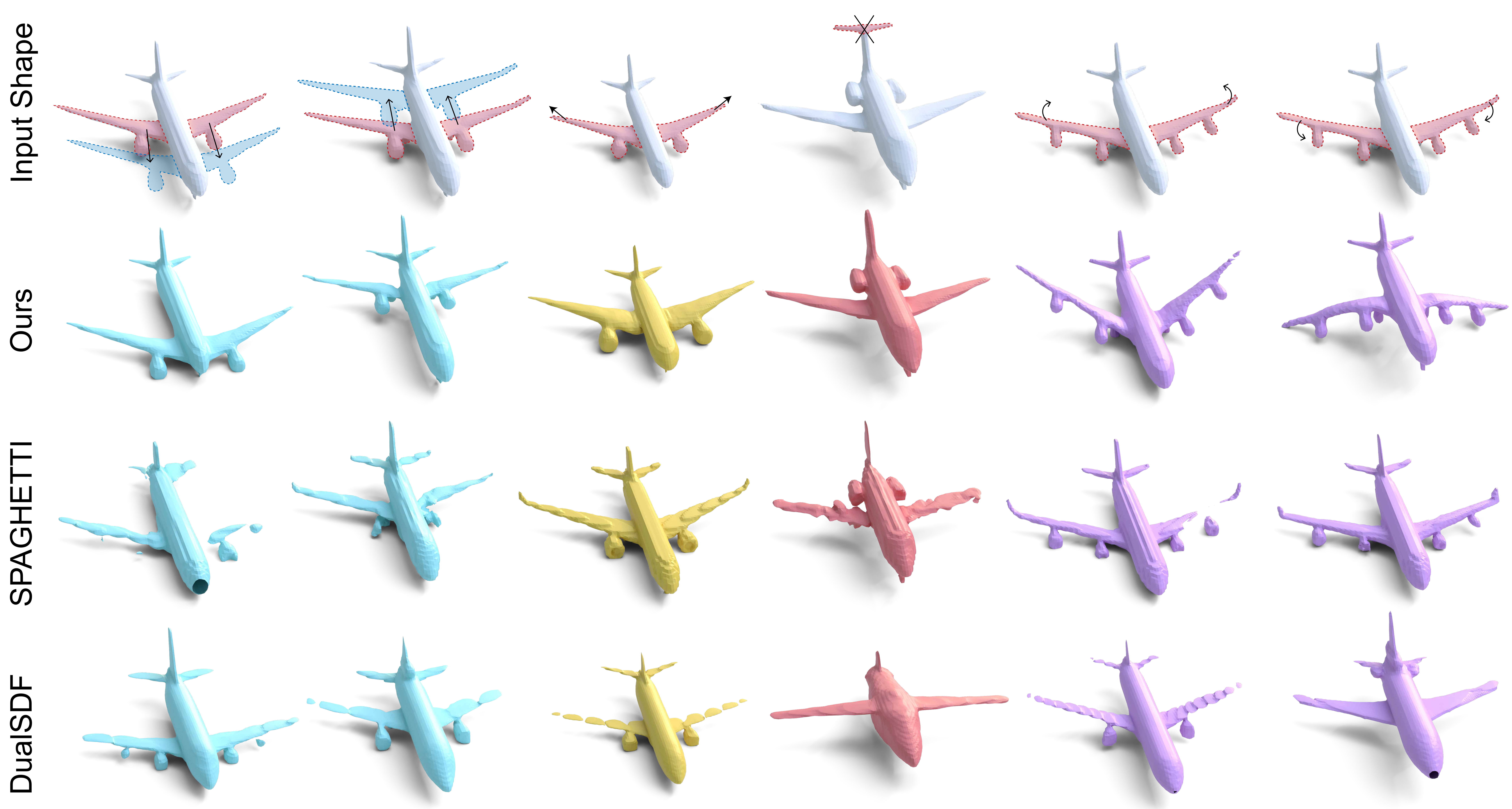}
    \caption{Editing comparisons on the airplane category using our method, DualSDF, and SPAGHETTI. Different colors indicate different types of editing operations.
    }
    \label{fig:exp-op-airplane}
\end{figure}

We categorize and visualize different editing operations to compare our method with SPAGHETTI and DualSDF. For chairs, representative results for each operation, including move, copy, delete, rotate, scale, and mix, are shown in Figures~\ref{fig:exp-op-copy} to~\ref{fig:exp-op-mix}. Since DualSDF’s interface relies exclusively on point-dragging deformations, it does not natively support explicit part deletion or mixing. Consequently, editing operations other than move and scale are only compared against SPAGHETTI. Additional visual comparisons for airplanes are presented in Figure~\ref{fig:exp-op-airplane}.

DualSDF offers ease of use through intuitive point dragging, automatically adjusting other parts based on the latent space. However, such edits operate on entangled latent codes, leading to the loss of original shape features and undesired structural changes. Often, edited shapes become unrecognizable compared to the source, and the lack of support for asymmetric editing limits its utility for fine-grained part manipulation.

SPAGHETTI, like our method, uses Gaussian ellipsoids to represent part poses and, in theory, can support the same set of editing operations. However, it directly encodes raw ellipsoid parameters instead of six representative endpoints and does not explicitly account for part-wise isolation. As a result, its editing outcomes suffer from interference. For example, when a part is moved significantly, it often remains anchored to its original location due to implicit biases, resulting in either part distortion or global warping. In copy operations, duplicate parts with identical identities tend to interfere with each other, leading to visual confusion. Rotation and scaling edits are also less effective, as pose information may be ignored or poorly disentangled from shape encoding.

In contrast, our method leverages attention-guiding part isolation and explicit pose encoding through ellipsoid vertices. This design ensures that each edit, whether a move, copy, or rotation, is interpreted as a direct and localized instruction, producing intuitive results without interference. Moreover, our model is not limited to shape-cutting operations. When edits create gaps or fractures, it learns to generate new bridging structures based on part relationships, yielding coherent and plausible reconstructions. Together, vertex-based pose encoding and attention regularization enable robust separation and accurate reconstruction of each edited part.

\begin{figure}[t]
    \centering
    \includegraphics[width=\linewidth]{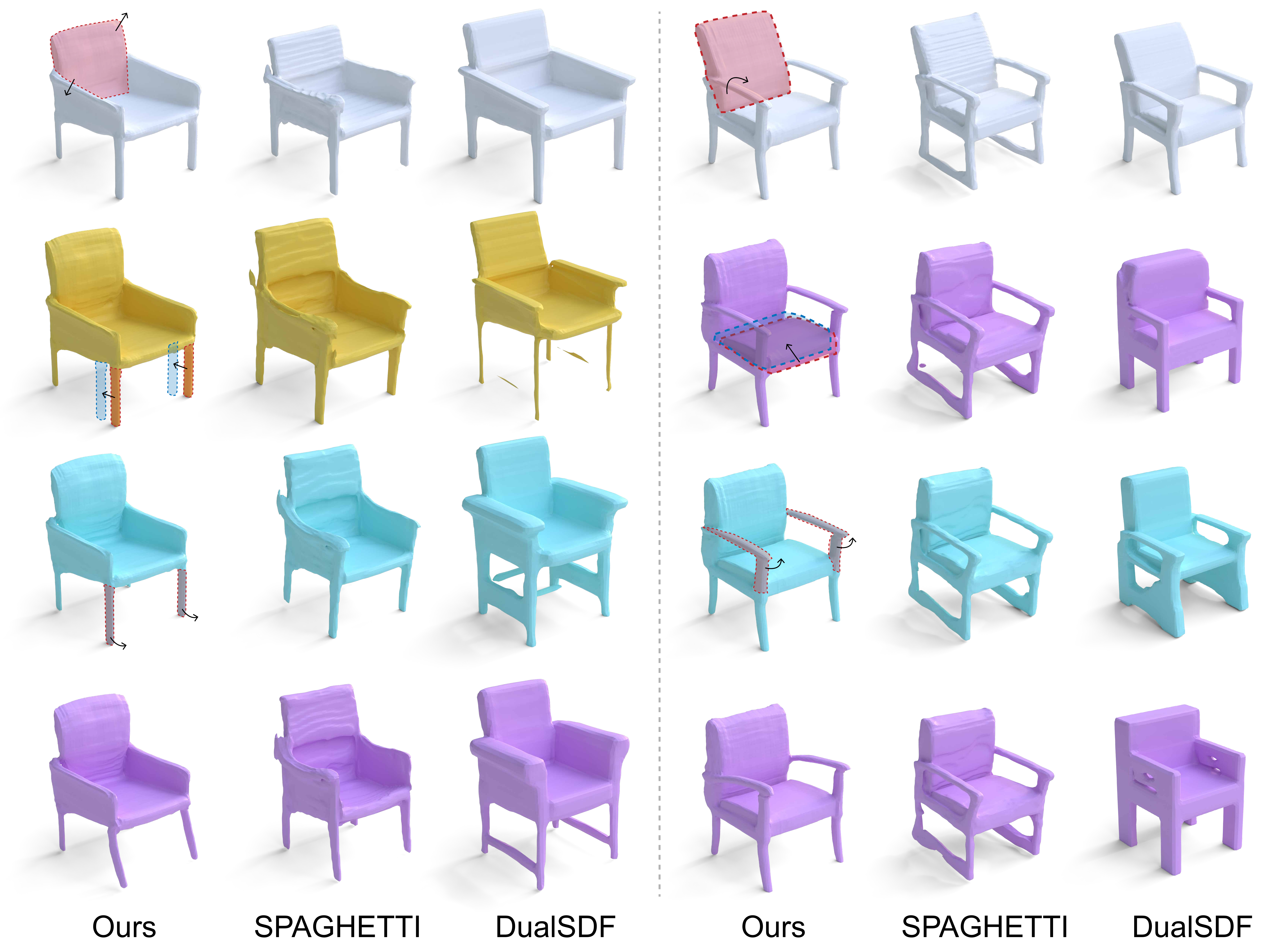}
    \caption{Comparison of sequential editing results. Each column shows the process of applying multiple edits to a shape. 
    }
    \label{fig:exp-continous-edit}
\end{figure}

\subsection{User Study}

\begin{table}[t]
  \centering
  \caption{Mean user ratings (on a scale of 1 to 5) for various editing operations performed using our method and baseline methods. ``-'' indicates that the user was unable to achieve the specified edit.}
  \label{tab:study}
    \begin{tabular}{lccc}
      \toprule
      \textbf{Operation} & \textbf{Ours} & \textbf{SPAGHETTI} & \textbf{DualSDF} \\
      \midrule
      Copy   & \textbf{4.07} & 2.18 & - \\
      Move   & \textbf{4.37} & 1.76 & 2.32 \\
      Delete & \textbf{4.37} & 2.21 & - \\
      Rotate & \textbf{3.92} & 2.86 & - \\
      Scale  & \textbf{3.89} & 2.40 & - \\
      Mix    & \textbf{4.13} & 2.06 & - \\
      \bottomrule
    \end{tabular}%
\end{table}

To evaluate editing performance in realistic scenarios, we conducted a user study. As preparation, we recruited a user with sufficient 3D editing experience to interact directly with each system. For every test shape, a specific editing goal, such as moving, rotating, scaling, or duplicating a part, was defined in advance. The user was free to perform any number of interactions until the desired edit was achieved or deemed impossible to reach. This setup ensures fairness across different editing interfaces by allowing users to explore the full potential of each method.

We then gathered feedback from 34 participants who were asked to rate the editing outcomes generated by different methods. Each participant evaluated a set of edited shapes across three criteria: (1) how well the result fulfilled the intended editing goal, (2) the naturalness and plausibility of the output, and (3) the preservation of key shape-specific features. As summarized in Table~\ref{tab:study}, our method consistently outperformed SPAGHETTI and DualSDF across all categories. For example, in the move operation, our method received an average score of 4.37 (out of 5), significantly higher than SPAGHETTI’s 1.76 and DualSDF’s 2.32. 
For more complex edits that are not natively supported by the baselines, such as rotation, scaling, or part copying, our approach demonstrated strong robustness and fidelity, confirming its effectiveness and versatility for part-based shape editing.

\subsection{Sequential Editing Robustness}

An editable shape representation must support not only single-step edits but also sequences of user operations. Figure~\ref{fig:exp-continous-edit} illustrates the results of applying multi-step editing sequences using DualSDF, SPAGHETTI, and our method.
With DualSDF, accumulated edits gradually degrade shape identity as latent features drift, eventually resulting in outputs that are no longer recognizable relative to the original shapes. SPAGHETTI suffers from increasing structural distortion due to the entangled part representations; successive edits compound interference effects, deteriorating overall shape quality.
In contrast, our method maintains stable reconstructions throughout the editing process. By treating each part as an independent entity and using explicit pose-aware decoding with attention-based guidance, our model ensures that parts preserve their original identity even after multiple edits. This robustness is essential for real-world applications, where users often require flexible, incremental modifications without compromising the integrity of the shape.

\subsection{Generation Ability}

\begin{figure}[t]
    \centering
    \includegraphics[width=\linewidth]{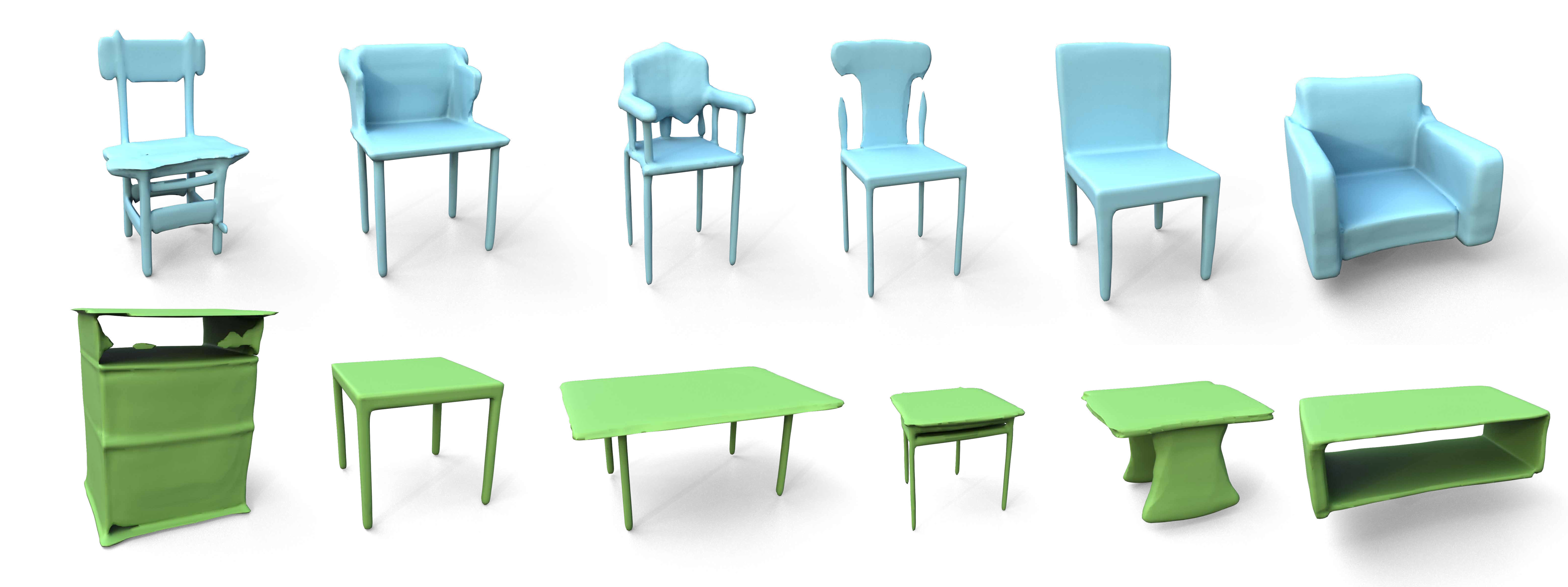}
    \caption{Random generation results produced by our diffusion model, showcasing diverse and plausible 3D shapes generated from latent representations.
    }
    \label{fig:random_generate}
\end{figure}

\begin{figure}[t]
    \centering
    \includegraphics[width=\linewidth]{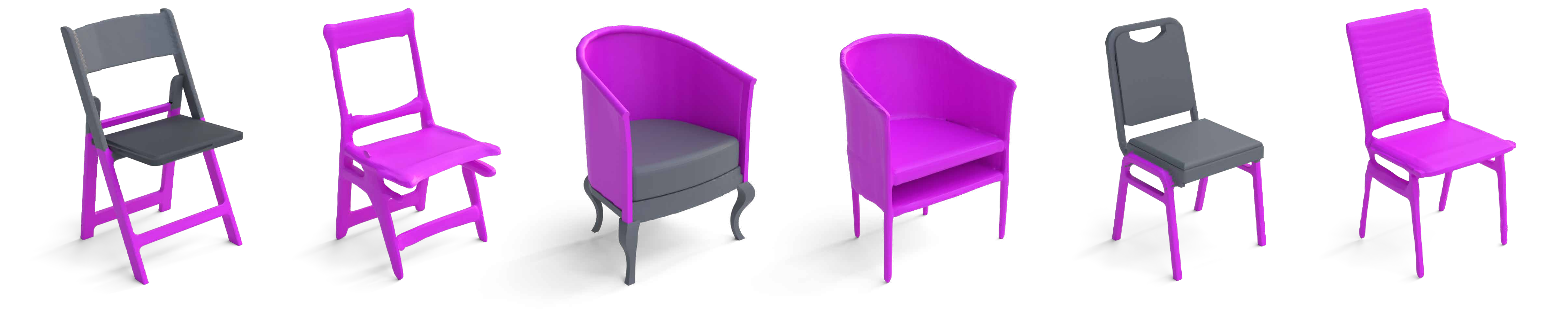}
    \caption{Shape completion results using our diffusion model, demonstrating the ability to generate coherent and natural completions for partially missing shapes.
    }
    \label{fig:complete_generate}
\end{figure}

\begin{figure*}[t]
    \centering
    \includegraphics[width=\linewidth]{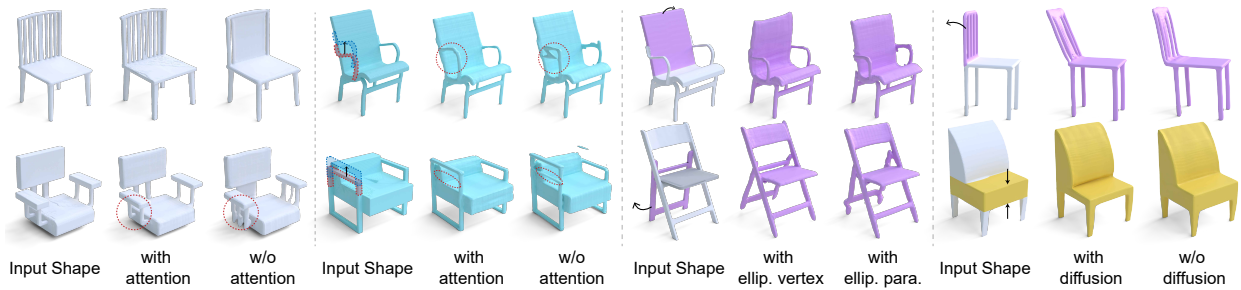}
    \caption{Ablation study on attention-guided decoding, Gaussian-ellipsoid representation, and diffusion-based post-edit refinement. From left to right: effect of the attention-guiding loss on reconstruction quality; comparison of asymmetric editing with and without attention guidance; effect of using six ellipsoid endpoints versus raw parameters for handling rotation edits; impact of applying a diffusion model to improve the plausibility of edited shapes.
    }
    \label{fig:ablation}
\end{figure*}

\begin{figure}
    \centering
    \includegraphics[width=\linewidth]{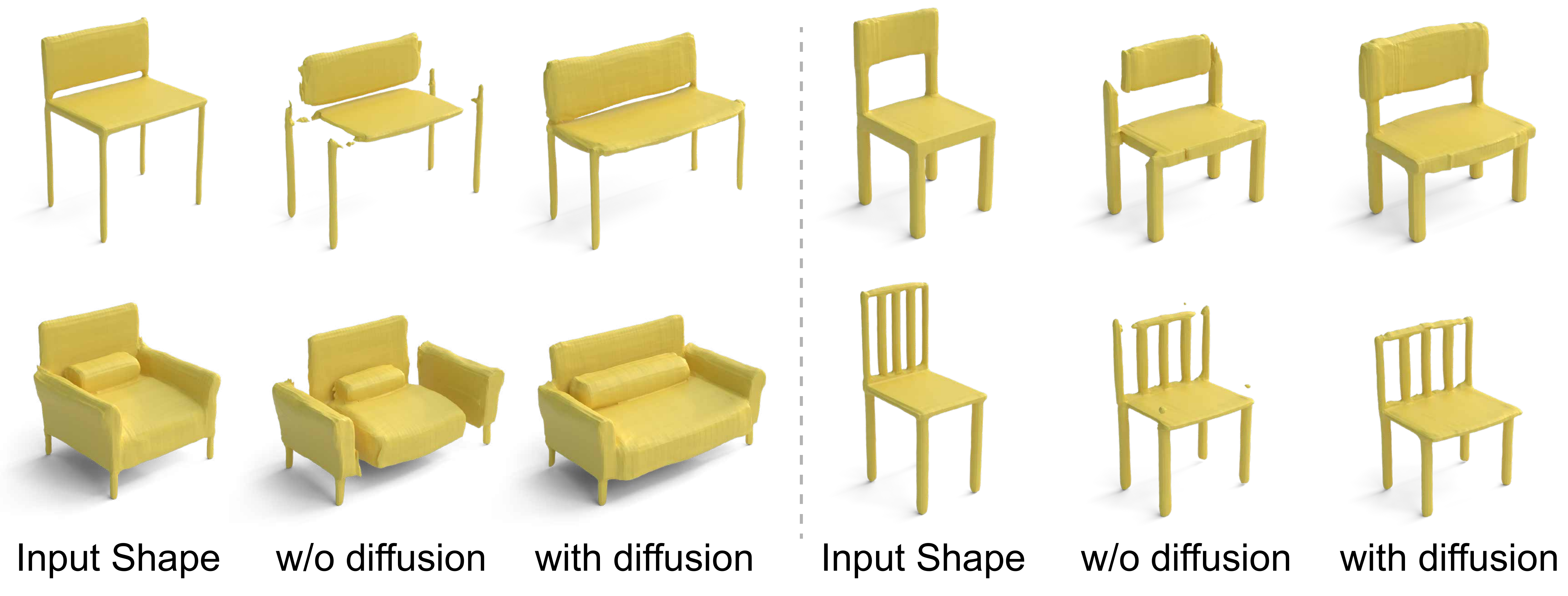}
    \caption{Qualitative results under extreme editing scenarios. Input shapes are randomly sampled and the positions of all parts are exaggeratedly shifted, for example by stretching them along the horizontal axis. This produces artifacts such as unnatural gaps, broken connections, and floating fragments. While the results without diffusion remain corrupted, our diffusion-based refinement yields coherent and structurally plausible reconstructions.}
    \label{fig:extreme-edit}
\end{figure}

By training a diffusion model on intermediate shape representations, we enable robust shape generation capabilities. While our method is primarily designed for editing rather than generation, it effectively handles the challenge of working with a variable number of parts without explicit identity information (e.g., distinguishing between a chair back and a seat). Despite this inherent complexity, we demonstrate the versatility of our method through two tasks: part random generation and shape completion.

Figures \ref{fig:random_generate} and \ref{fig:complete_generate} illustrate examples of shapes generated randomly and via completion, respectively. These shapes are represented using our intermediate shape representation, enabling further edits and subsequent Global Semantic Adjustment. This capability supports an interactive workflow that allows users to iteratively edit and refine generated shapes to meet their specific needs.

\subsection{Ablation Study}\label{sec:Ablation}

To evaluate the effectiveness of our ellipsoid vertex encoding and attention-based decoding, we conduct a series of ablation experiments and present representative results in Figure~\ref{fig:ablation}.

In Figure~\ref{fig:ablation}, the first and second sets of columns demonstrate the impact of attention guidance on distinguishing spatially adjacent parts, such as the multiple slats on a chair back. 
Specifically, the first set of columns demonstrates the reconstruction results with and without attention guidance, and the second set of columns shows the editing results.
Without attention guidance, the model tends to merge spatially adjacent parts, resulting in blurred or fused outputs. The attention-guiding loss also plays a key role in preventing information leakage from correlated (particularly symmetric) parts. This ensures that asymmetric edits produce localized changes without unintentionally affecting unrelated regions.

The third set of columns of Figure~\ref{fig:ablation} shows the benefit of using ellipsoid vertex encoding. This representation significantly improves the model's sensitivity to rotation and scaling edits. For instance, in vertically elongated parts, the top and bottom vertices serve as clear geometric anchors, helping the model better understand pose adjustments. Without this encoding, pose-related signals, such as orientation and scale, are often overlooked or absorbed into the shape code.

The fourth set of columns of Figure~\ref{fig:ablation} shows the utility of incorporating the diffusion model. In the case of large edits, such as rotations, scalings, or translations, directly applying these operations can yield shapes that are structurally implausible or visually inconsistent due to mismatches between the updated pose and the original encoding. By refining part embeddings under the new pose via diffusion, the model produces more coherent and plausible results, enhancing robustness in complex editing scenarios.

\subsection{Additional Evaluation under Extreme Editing}\label{sec:extreme}

To further validate the effectiveness of our diffusion module in repairing artifacts caused by large-scale shape edits, we conducted a controlled experiment. Specifically, we randomly sampled shapes from the dataset and applied extreme geometric modifications by stretching the positions of all parts along the horizontal axis and scaling them in other directions. These manipulations resulted in unnatural deformations, structural breakages, and floating fragments, which simulate the kind of severe defects introduced by aggressive edits.

In order to prevent the model from simply restoring the original shape, we fixed the position of each part and applied our diffusion-based denoising process to repair the corrupted geometry. As illustrated in Figure~\ref{fig:extreme-edit}, the results without diffusion exhibit noticeable defects such as disjointed parts and distorted surfaces. In contrast, the shapes refined with the diffusion module are visually coherent, structurally consistent, and free from floating artifacts. This demonstrates the ability of our approach to effectively restore plausibility even under challenging editing conditions.

\section{Conclusion}

We present a novel framework for editable 3D shape representation that enables intuitive and semantically consistent part-level manipulation. By decoupling shape and pose through latent encoding and Gaussian ellipsoid parameterization, our method allows users to freely transform individual parts without unintended interference. A vertex-based ellipsoid encoding ensures accurate interpretation of spatial transformations, while an attention-based decoding process preserves semantic locality and minimizes information leakage between adjacent or symmetric parts.

Extensive experiments validate our method's effectiveness across reconstruction, single-step editing, sequential editing, and generalization to various categories. Compared to prior work, our method achieves competitive reconstruction quality while significantly improving editing flexibility and robustness. Ablation studies further confirm the importance of our design choices, including vertex-based pose representation, attention regularization, and optional diffusion refinement.

Overall, our system provides a practical foundation for part-aware 3D editing, supporting both precise direct manipulation and high-level structural refinement. One limitation of our method, however, is that it treats parts as the smallest editable units. This restricts finer-grained control, such as modifying sub-part geometry or adjusting local details within a part. Future work may explore hierarchical representations that allow editing at multiple granularities or introduce optimization strategies that adaptively refine part-level shape embeddings to better satisfy complex user intentions.

{\small
\bibliographystyle{eg-alpha-doi} 
\bibliography{main} 
}

\clearpage
\appendix
\onecolumn
\section*{Supplementary Material}

\section{Implementation Details}

\textbf{Part Feature Extraction.} 
Each part is first converted into a binary voxel grid by applying Poisson Surface Reconstruction (PSR) to a ShapeNet point cloud and thresholding the resulting SDF values. This binary representation effectively masks connectivity information at cut boundaries: SDF-based inputs would retain high distance values across cuts, biasing the model toward original connections.

We then transform the dense grid into a sparse voxel representation and apply a sequence of 3×3×3 convolutions and max-pooling layers. Feature channels increase from 1 → 8, doubling at each pooling until reaching 512 channels. The network branches into two heads: a 10-dimensional pose head (for Gaussian parameters) and a 512-dimensional shape code head. Both heads are implemented with 1×1×1 convolutions.

\textbf{Pose Feature Mixing.} 
For each edited Gaussian proxy, we extract the six ellipsoid endpoints along its principal axes at radius 1.75. We encode these 18 coordinates with a SIREN-based positional mapping $\phi(\cdot)$, then concatenate the result with the 512-dimensional shape code and pass through a four-layer MLP (with skip connections and layer normalization) to produce the 256-dimensional part embedding. We adopt random affine perturbations during training: translations sampled uniformly in [–0.5, 0.5], scales as $\exp(\mathcal{U}(-0.5,0.5))$, and rotations in ($-\pi/4$, $\pi/4$) applied with 10\% probability to avoid ambiguity in principal axis orientation.

\textbf{Attention-based Shape Decoding.} 
A query point $\mathbf{x}\in\mathbb{R}^3$ is first encoded by the same SIREN mapping $\phi(\mathbf{x})$. We then use a transformer decoder without self-attention, so that points remain independent, to attend over the set of part embeddings and produce a 128-dimensional local feature $\mathbf{h}_x$. Concatenating $\phi(\mathbf{x})$ and $\mathbf{h}_x$, we decode occupancy via a NeRF-style 4-layer MLP.

\textbf{Diffusion model.} 
We trained the diffusion module following the architectural and training configurations of 3DShape2VecSet~\cite{Zhang20233DShape2VecSet} and Improved Denoising Diffusion Probabilistic Models~\cite{Nichol2021ImprovedDD}. 

To stabilize optimization, we applied rescaling to different components of the concatenated input vector ($1+15+256$ dimensions) in order to balance their relative magnitudes. The binary validity flag was scaled into the range $[-0.01,0.01]$, ensuring that it is only determined at the final stage without interfering with the denoising process. For the Gaussian embedding $\mathbf{g}_i$, we multiplied the positional and ellipsoidal radius components by 5, while keeping the rotation matrix unchanged, since the unit-norm vectors of the rotation matrix naturally span a larger range. We further applied a random scaling factor between $0.8$ and $1.2$ to the input shapes to improve robustness. For the shape code $\mathbf{z}_i$, we applied a scale of $0.1$, which keeps $\mathbf{z}_i$ relatively blurred during the denoising of $\mathbf{g}_i$. This prevents the model from overfitting to a fixed correspondence between $\mathbf{g}_i$ and $\mathbf{z}_i$, and we found this adjustment to be crucial for achieving high-quality results. All rescaled values are reverted to their original ranges before being fed into the decoder.

We used the standard 1000-step diffusion schedule, where generating a complete set of parts takes approximately 5 seconds. For refinement tasks, we adopt a shortened sequence consisting of 100 noising steps followed by 100 denoising steps.

\textbf{Training and Inference.} 
During training, we sample 3,000 points uniformly inside the shape’s bounding box and add Gaussian noise ($\sigma=0.02$) around the surface, then resample another 3,000 points to balance interior/exterior supervision. At inference time, we generate a 128³ grid of query points and process them in batches of 1,000,000 through the decoding network, computing occupancy values in approximately 1 s per shape. Performance can be further improved by coarse-to-fine sampling strategies around predicted surfaces.

\section{More Random Generation Result} 

\begin{figure*}
    \centering
    \includegraphics[height=0.95\textheight]{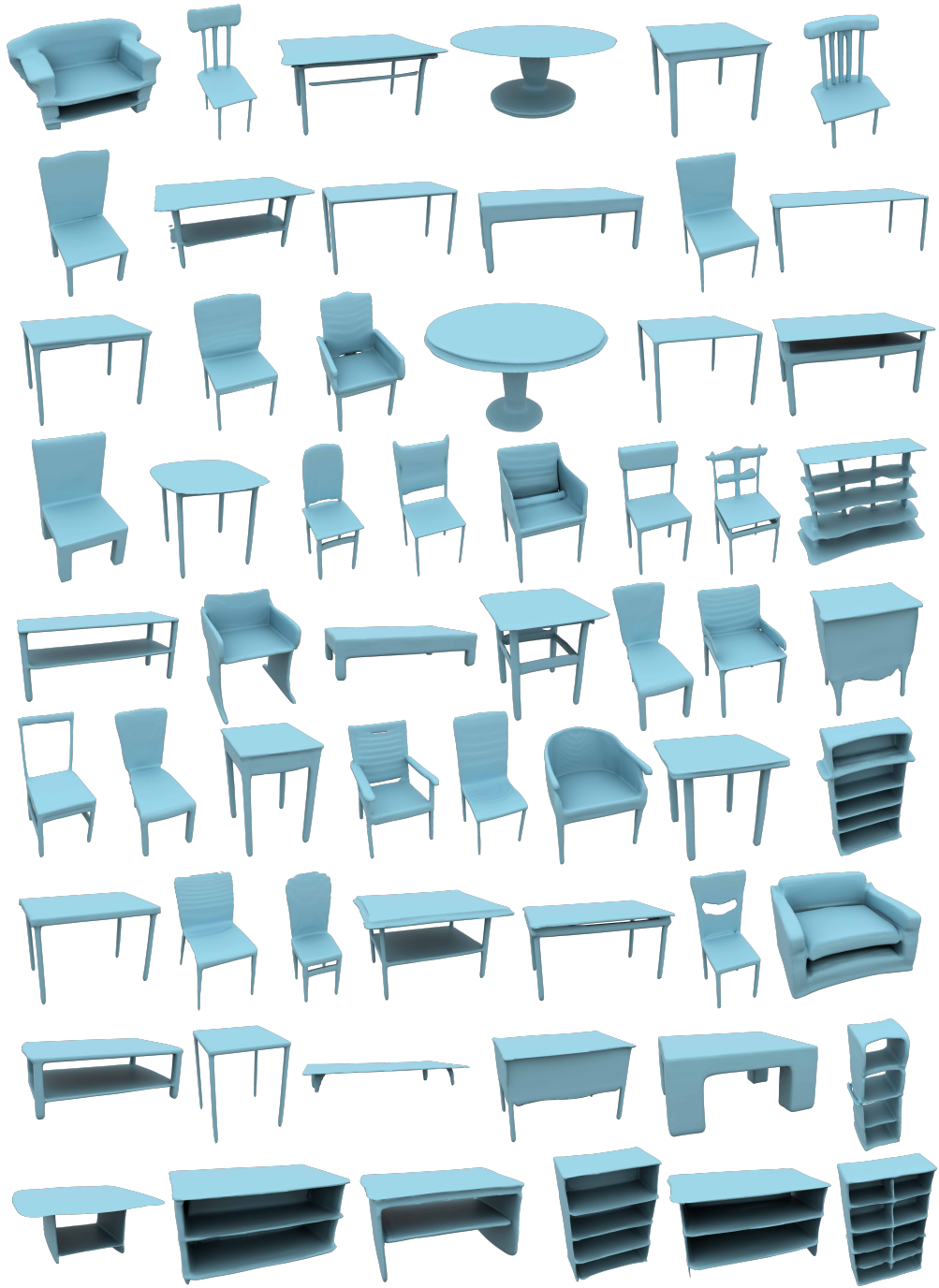}
    \caption{More visual results of shape random generation.}
    \label{fig:exp-random-gen-supp}
\end{figure*}

Due to space limitations, we present only a small portion of the randomly generated results in the main text. In Fig. \ref{fig:exp-random-gen-supp}, additional generated results are provided. The shapes produced through our shape representation training exhibit significant structural diversity.

\section{Structure-Preserving Shape Generation} 

\begin{figure*}
    \centering
    \includegraphics[width=\linewidth]{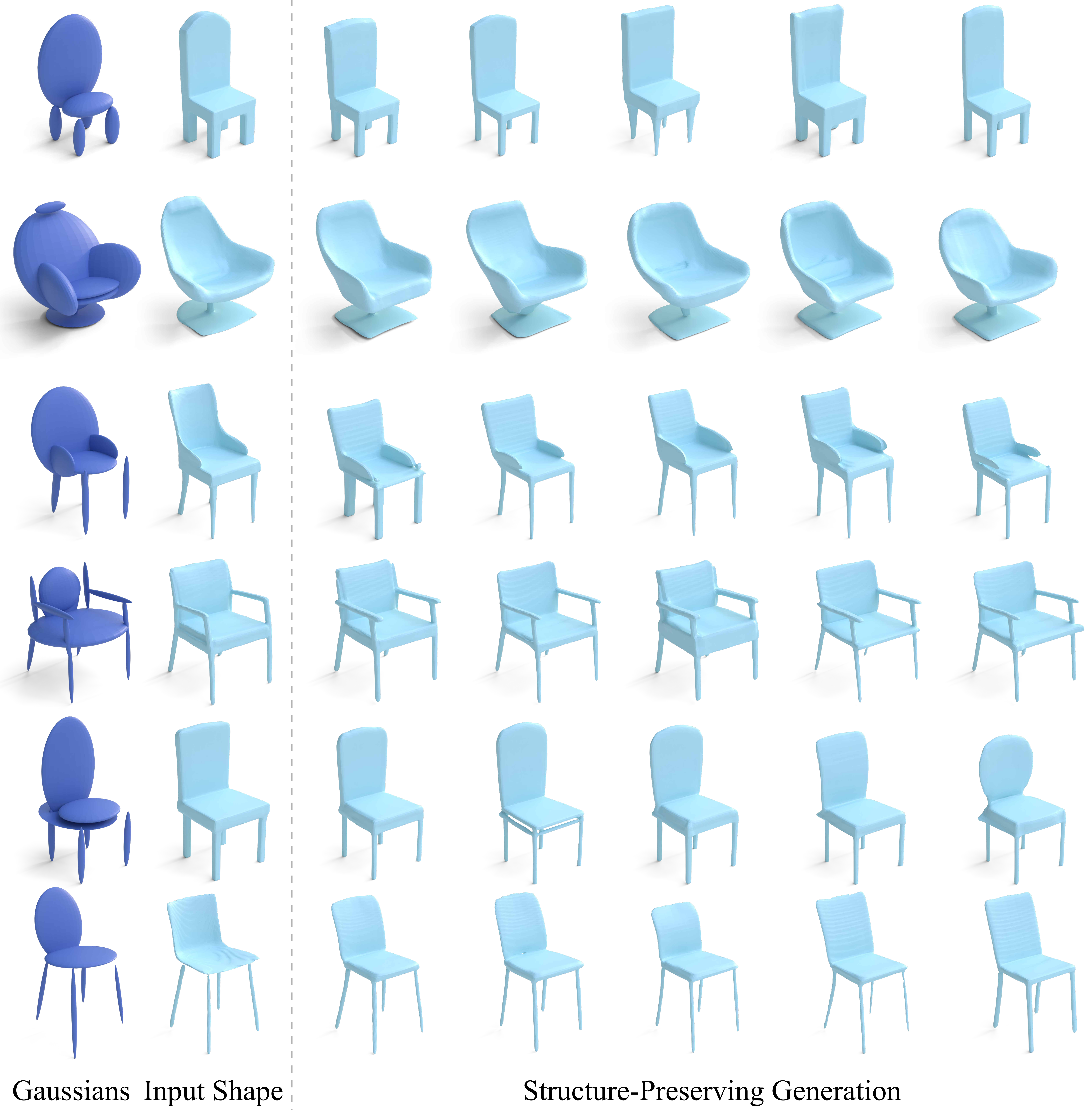}
    \caption{Visual results of structure-preserving shape generation.}
    \label{fig:exp-structure-fix-gen}
\end{figure*}

Our diffusion generative model is capable of generating a wide variety of shapes conditioned on a given structure. As illustrated in Fig. \ref{fig:exp-structure-fix-gen}, several examples of such results are presented. Additionally, we provide a preview of the corresponding part-structure representation $\mathbf{g}$ for each generated shape.

\section{Editing and Mixing Shapes from Scanned Objects} 

\begin{figure*}
    \centering
    \includegraphics[width=\linewidth]{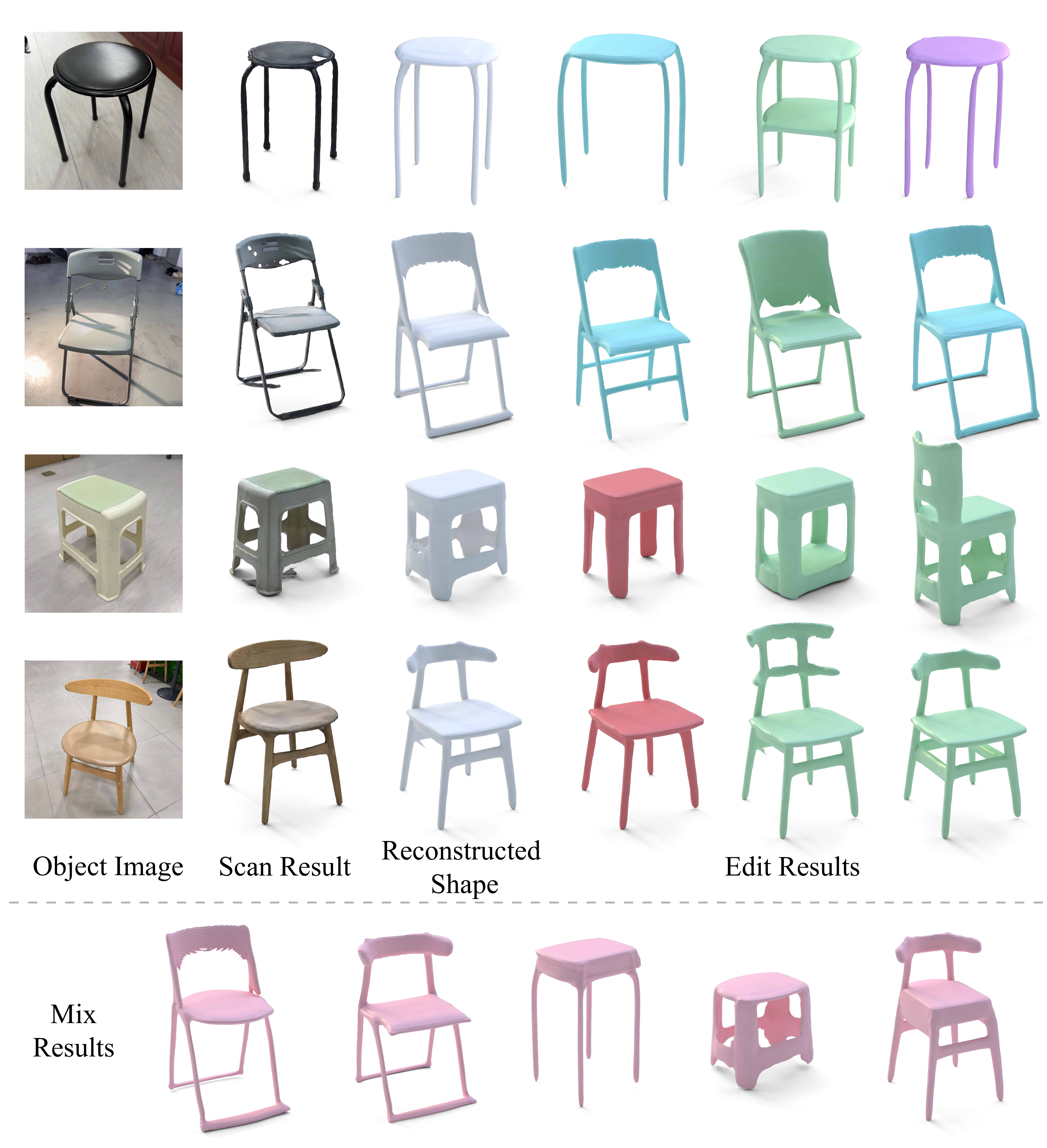}
    \caption{Visual results of editing and mixing shapes from scanned objects.}
    \label{fig:exp-scanning}
\end{figure*}

We utilized the Apple iPhone 16 Pro to scan several types of chairs. Our method requires only a few annotations of bounding boxes to import scanned real-world shapes. As shown in Fig. \ref{fig:exp-scanning}, our approach effectively demonstrates the editing of real-world shapes. Additionally, it can address and partially rectify imperfections in scanning.

\end{document}